\documentclass[useAMS,usenatbib]{mn2e}
\usepackage{verbatim} 
\usepackage{natbib} 
\usepackage{amsmath} 
\usepackage{amsbsy}
\usepackage{amssymb}
\usepackage{mathrsfs} 
\usepackage{lscape} 
\usepackage{graphicx}
\usepackage{epstopdf}
\usepackage{deluxetable} 
\usepackage{fixltx2e} 
\usepackage{ctable}

\newcommand{\lya}{Ly$\alpha$}

\newcommand{\ion}[2]{#1\,{\small{#2}}}
\newcommand{\hi}{\ion{H}{I}}

\newcommand{\ciii}{\ion{C}{III}}
\newcommand{\civ}{\ion{C}{IV}}

\newcommand{\nv}{\ion{N}{V}}

\newcommand{\ovi}{\ion{O}{VI}}

\newcommand{\neviii}{\ion{Ne}{VIII}}
\newcommand{\siiii}{\ion{Si}{III}}
\newcommand{\siiv}{\ion{Si}{IV}}

\newcommand{\cloudy}{{\sc CLOUDY}}

\title[UV Emission from the CGM of High-z Galaxies]{Strongly Time-Variable Ultra-Violet Metal Line Emission from the Circum-Galactic Medium of High-Redshift Galaxies}

\author[Sravan et al.]{Niharika Sravan$^1$\thanks{niharika@u.northwestern.edu}, Claude-Andr\'e Faucher-Gigu\`ere$^{1}$, Freeke van de Voort$^{2,3}$,\newauthor Du\v{s}an Kere\v{s}$^{4}$,  Alexander L. Muratov$^{4}$, Philip F. Hopkins$^{5}$, Robert Feldmann$^{2}$,\newauthor Eliot Quataert$^{2}$, and Norman Murray$^{6,7}$\\ 
$^{1}$Center for Interdisciplinary Exploration and Research in Astrophysics (CIERA), and\\
Department of Physics and Astronomy, Northwestern University, 2145 Sheridan Road, Evanston IL 60208, USA
\\
$^{2}$Department of Astronomy and Theoretical Astrophysics Center, University of California, Berkeley, CA 94720-3411, USA
\\
$^{3}$Academia Sinica Institute of Astronomy and Astrophysics, PO Box 23-141, Taipei 10617, Taiwan
\\
$^{4}$Department of Physics, Center for Astrophysics and Space Science, University of California, San Diego, 9500 Gilman Drive,\\ La Jolla, CA 9209, USA
\\
$^{5}$TAPIR, Mailcode 350-17, California Institute of Technology, Pasadena, CA 91125, USA
\\
$^{6}$Canadian Institute for Theoretical Astrophysics, 60 St. George Street, University of Toronto, ON M5S 3H8, Canada
\\
$^{7}$Canada Research Chair in Astrophysics
}

\begin{document}
\maketitle


\begin{abstract}
We use cosmological simulations from the Feedback In Realistic Environments (FIRE) project, which implement a comprehensive set of stellar feedback processes, to study ultra-violet (UV) metal line emission from the circum-galactic medium of high-redshift ($z=2-4$) galaxies. 
Our simulations cover the halo mass range $M_{\rm h} \sim 2\times10^{11} - 8.5\times10^{12}$ M$_\odot$ at $z = 2$, representative of Lyman break galaxies. 
Of the transitions we analyze, the low-ionization C III (977 \AA) and Si III (1207 \AA) emission lines are the most luminous, with C IV (1548 \AA) and Si IV (1394 \AA) also showing interesting spatially-extended structures.  
The more massive halos are on average more UV-luminous. 
The UV metal line emission from galactic halos in our simulations arises primarily from collisionally ionized gas and is strongly time variable, with peak-to-trough variations of up to $\sim2$ dex. 
The peaks of UV metal line luminosity correspond closely to massive and energetic mass outflow events, which follow bursts of star formation and inject sufficient energy into galactic halos to power the metal line emission. 
The strong time variability implies that even some relatively low-mass halos may be detectable. 
Conversely, flux-limited samples will be biased toward halos whose central galaxy has recently experienced a strong burst of star formation. 
Spatially-extended UV metal line emission around high-redshift galaxies should be detectable by current and upcoming integral field spectrographs such as the Multi Unit Spectroscopic Explorer (MUSE) on the Very Large Telescope and Keck Cosmic Web Imager (KCWI).
\end{abstract}


\section{Introduction}
In the cosmic web, galaxies form in dark matter over-densities. 
In order to sustain star formation across cosmic time, they continuously accrete their baryonic fuel from the surrounding intergalactic medium along with dark matter \citep[IGM;][]{kk05, pw09, blm10, fkm11}. 
Some of this gas is later returned to the circum-galactic medium (CGM), along with metals produced in stars, via powerful galactic winds driven by feedback from stars \citep[e.g.][]{ahs01, sh03, od06, mse10, ses10, 2010ApJ...719.1503R} and active galactic nuclei \citep[AGN; e.g.,][]{fbt10, wst11}. 
Since the inflows and outflows that regulate galaxy formation are mediated through the CGM, 
measurements of the gas properties in the CGM are a powerful approach to understanding galaxy assembly and evolution.

To date, most of our observational constraints on the physical state of the CGM have come from absorption line studies. 
Absorption lines have the advantage that even low-density, faint gas can be probed if there is a suitable background source. 
Absorption line measurements however have several imitations. 
Firstly, suitable background sources (typically, quasars) are rare so that there is typically only one (or no) background sight line per foreground galaxy \citep[e.g.,][]{2005ApJ...629..636A, 2009ApJ...690.1558P, 2012ApJ...750...67R, 2013ApJ...777...59T}. 
Because the area covered scales as the square of the impact parameter, the resulting measurements preferentially probe large impact parameters from foreground galaxies. 
Finally, because of the 1D nature of absorption line constraints it is often unclear what physical structures are probed.\footnote{With the advent of 30-m class optical telescopes, it will become possible to sample galaxy halos more densely by using ordinary galaxies as background sources \citep[e.g.,][]{ses10}. With sufficiently dense sampling, absorption line measurements can produce 3D tomographic maps of the foreground gas.}

Integral field measurements of emission lines are complementary to absorption line measurements.
In particular, they allow us to directly obtain a 3D picture of the distribution of gas in the CGM by combining the 2D map of emission on the sky with line-of-sight velocity information. 
In recent years, we have indeed learned a great deal about spatially-resolved kinematics of high-redshift galaxies using this technique \citep[e.g.,][]{2008ApJ...687...59G, 2009ApJ...706.1364F, 2009ApJ...699..421W, 2009ApJ...697.2057L}. 
Additionally, since gas emissivity scales with the square of density, emission line studies preferentially probe the dense gas closer to galaxies. 
To date, it has generally not been possible to make 3D emission line maps of the CGM of galaxies because halo gas is much fainter than galactic gas, thus pushing the capabilities of existing astronomical instruments. 
The situation is however changing thanks to efforts to build sensitive integral field spectrographs (IFS) that are well suited to detect the low surface brightness features expected from inflows and outflows around galaxies.

Thanks to the abundance of hydrogen, \hi\ Lyman-$\alpha$ (Ly$\alpha$) is the most easily detectable line from the CGM. 
Our first glimpses of the CGM emission have come from spatially-extended Ly$\alpha$ sources known as ``Ly$\alpha$ blobs'' (LABs).  
The blobs have line luminosities up to $\sim 10^{44}$ erg s$^{-1}$ and spatial extents sometimes exceeding $100$ proper kpc \citep[][]{2000ApJ...532..170S, 2004AJ....128..569M, 2009ApJ...693.1579Y}. 
The physical nature of LABs is not yet well understood but there is growing evidence that they are often powered by an energetic source such as an AGN or a starburst galaxy \citep[which can be beamed away from the line of sight and appear obscured;][]{2009ApJ...700....1G, 2014ApJ...793...22G, 2015arXiv150105312P}. 
AGN and star-forming galaxies can induce the CGM to glow through several mechanisms. 
Ly$\alpha$ photons can be produced inside galaxies by the processing of ionizing photons that are absorbed by the interstellar medium (ISM) and scatter in a diffuse halo as they escape the galaxy. 
Diffuse Ly$\alpha$ halos are now inferred to be generically produced by ordinary star-forming galaxies by this mechanism \citep[][]{2011ApJ...736..160S}. 
Ionizing photons that escape galaxies but are absorbed in the CGM can also produce fluorescent Ly$\alpha$ emission \citep[e.g.,][]{1996ApJ...468..462G, cpl05, 2010ApJ...708.1048K}. 
Alternatively, energy can be injected in the CGM as galactic winds driven by stellar or AGN feedback encounter halo gas. This energy can then power \lya\ emission from the CGM \citep[][]{2000ApJ...532L..13T, 2001ApJ...562L..15T}. 
We indeed show in this paper that galactic winds can power significant metal line emission from circum-galactic gas, and that this process induces dramatic time variability. 
Finally, spatially-extended Ly$\alpha$ emission can be powered by gravitational potential energy that is dissipated as gas falls into dark matter halos \citep[][]{2000ApJ...537L...5H, fkg01, dl09, 2010MNRAS.407..613G, 2012MNRAS.423..344R}. 
Calculations that take into account self-shielding of dense gas suggest however that gravity-driven Ly$\alpha$ emission is in general too faint to explain the most luminous LABs \citep[][]{fkd10}. 

\begin{footnotesize}
\ctable[caption={{\normalsize Parameters of the Simulations Analyzed in this Work}},center]{lcccccl}{ 
\label{t:simulations}
\tnote[ ]{Parameters describing the initial conditions for our simulations (units are physical): \\
{\bf (1)} Name: Simulation designation. Simulations ${\bf mx}$ have a main halo mass $\sim10^{x} M_{\odot}$ at $z=0$.\\
{\bf (2)} $M_{\rm h}^{z=2}$: Mass of the main halo at $z=2$. \\
{\bf (3)} $m_{\rm b}$: Initial baryonic (gas and star) particle mass in the high-resolution region. \\
{\bf (4)} $\epsilon_{\rm b}$: Minimum baryonic force softening (fixed in physical units past $z\sim10$; minimum SPH smoothing lengths are comparable or smaller). Force softening lengths are adaptive (mass resolution is fixed).\\
}
}{
\hline\hline
\multicolumn{1}{l}{Name} &
\multicolumn{1}{c}{$M_{\rm h}^{z=2}$} &
\multicolumn{1}{c}{$m_{\rm b}$} & 
\multicolumn{1}{c}{$\epsilon_{\rm b}$} & 
\multicolumn{1}{c}{$m_{\rm dm}$} & 
\multicolumn{1}{c}{$\epsilon_{\rm dm}$} \\ 
\multicolumn{1}{l}{\ } &
\multicolumn{1}{c}{$M_{\sun}$} &
\multicolumn{1}{c}{$M_{\sun}$} &
\multicolumn{1}{c}{pc} &
\multicolumn{1}{c}{$M_{\sun}$} &
\multicolumn{1}{c}{pc} \\
\hline
{\bf m12v} & 2.0$\times 10^{11}$ & 3.9$\times 10^{4}$ & 10 & 2.0$\times 10^{5}$ & 140 \\ 
{\bf m12q} & 5.1$\times 10^{11}$ & 7.1$\times 10^{3}$ & 10 & 2.8$\times 10^{5}$ & 140 \\ 
{\bf m12i}  & 2.7$\times 10^{11}$ & 5.0$\times 10^{4}$ & 14 & 2.8$\times 10^{5}$ & 140 \\ 
{\bf m13}  & 8.7$\times 10^{11}$ & 3.7$\times 10^{5}$ & 20 & 2.3$\times 10^{6}$ & 210 \\
\hline
{\bf z2h830} & 5.4$\times 10^{11}$ & 5.9$\times 10^{4}$ & 9 & 2.9$\times 10^{5}$ & 143 \\
{\bf z2h650} & 4.0$\times 10^{11}$ & 5.9$\times 10^{4}$ & 9 & 2.9$\times 10^{5}$ & 143 \\
{\bf z2h600} & 6.7$\times 10^{11}$ & 5.9$\times 10^{4}$ & 9 & 2.9$\times 10^{5}$ & 143 \\
{\bf z2h550} & 1.9$\times 10^{11}$ & 5.9$\times 10^{4}$ & 9 & 2.9$\times 10^{5}$ & 143 \\
{\bf z2h506} & 1.2$\times 10^{12}$ & 5.9$\times 10^{4}$ & 9 & 2.9$\times 10^{5}$ & 143 \\
{\bf z2h450} & 8.7$\times 10^{11}$ & 5.9$\times 10^{4}$ & 9 & 2.9$\times 10^{5}$ & 143 \\
{\bf z2h400} & 7.9$\times 10^{11}$ & 5.9$\times 10^{4}$ & 9 & 2.9$\times 10^{5}$ & 143 \\
{\bf z2h350} & 7.9$\times 10^{11}$ & 5.9$\times 10^{4}$ & 9 & 2.9$\times 10^{5}$ & 143 \\
\hline
{\bf MFz2\_B1} & 8.5$\times 10^{12}$ & 3.3$\times 10^{4}$ & 9 & 1.7$\times 10^{5}$ & 143 \\
\hline\hline
}
\end{footnotesize} 

In spite of a growing body of high-quality observations of spatially extended Ly$\alpha$ at high-redshift \citep[e.g.,][]{2008ApJ...681..856R, 2009ApJ...693L..49H, 2014ApJ...786..106M, 2014Natur.506...63C}, these observations have proved difficult to interpret because of the strong radiative transfer effects experienced by Ly$\alpha$ photons as they are scattered by interstellar and circum-galactic gas \citep[e.g.,][]{dhs06, vsm06, fkd10}. In effect, scattering scrambles the photons both spatially and spectrally, making it challenging to reliably infer the geometry and kinematics of the emitting gas. 

Emission from metal lines in the CGM can be powered by the same physical processes as Ly$\alpha$. 
Because metals are not as abundant, metal line emission is generally fainter and more difficult to detect than Ly$\alpha$. 
Observing metal lines however provides valuable complementary information, and in some respects provide a more direct window into the physics of circum-galactic gas flows. 
Most metal lines are optically thin and therefore not subject to significant photon scattering, unlike Ly$\alpha$. 
Furthermore, different metal ions probe different temperature regimes, allowing us to construct a more complete physical picture. 
In particular, ions with high ionization potential preferentially probe more diffuse, volume-filling gas whereas Ly$\alpha$ emission tends to peak around cold filaments \citep[e.g.,][]{2010MNRAS.407..613G, fkd10, 2012MNRAS.423..344R}. 
Since much of the cooling of the diffuse Universe occurs through rest-frame ultra-violet (UV) emission \citep[e.g.,][]{bas13}, rest-UV metal lines offer a powerful way to probe galaxy formation. 
Rest-frame UV from $z\sim2-4$, covering the peak of the cosmic star formation history \citep[e.g.,][]{2007ApJ...670..928B}, also has the benefit of being redshifted into the optical and thus is accessible using large optical telescopes. 

A number of optical integral field spectrographs (IFS) with the capacity to detect low surface brightness, redshifted rest-UV CGM emission have recently been commissioned or are planned for the near future. 
The Cosmic Web Imager \citep[CWI;][]{2010SPIE.7735E..24M} started taking data on the Hale 200'' telescope at Palomar Observatory in 2009 and the first science results on luminous spatially-extended Ly$\alpha$ sources at $z\sim2-3$ have recently been reported \citep[][]{mcm14a, mcm14b}. 
Its successor, the Keck Cosmic Web Imager\footnote{http://www.srl.caltech.edu/sal/keck-cosmic-web-imager.html} \citep[KCWI,][]{2010SPIE.7735E..21M}, to be mounted on the Keck II telescope at the W. M. Keck Observatory on Mauna Kea, is currently being developed. 
The Multi Unit Spectroscopic Explorer\footnote{http://muse.univ-lyon1.fr} \citep[MUSE;][]{2010SPIE.7735E...7B} on the Very Large Telescope (VLT) completed its commissioning in August 2014 and early science results are being reported \citep[e.g.,][]{2014MNRAS.445.4335F, 2015A&A...575A..75B, 2015MNRAS.446L..16R, 2015arXiv150905143W}. 
MUSE combines a wide field of view, excellent spatial resolution, and a large spectral range. 
These IFSs not only provide kinematic information not available with narrowband imaging but also enable more accurate background subtraction to characterize low surface brightness features.

Cosmological simulations have previously been used to predict UV emission from the IGM and the CGM \citep[e.g.,][]{fss04, bsb10, frv12, bs12, vs13}. 
Our simulations, from the FIRE (Feedback In Realistic Environments) project,\footnote{See the FIRE project web site at: http://fire.northwestern.edu.} improve on these previous analyses in several respects. 
Most of our zoom-in simulations have a gas particle mass of a few times $10^{4}$ M$_{\odot}$ and a minimum (adaptive) gas gravitational softening length of $\sim10$ proper pc. 
For comparison, the highest resolution simulations from the OWLS project analyzed by \cite{vs13} have a gas particle mass of $2.1\times10^{6}$ M$_{\odot}$ and a maximum gas gravitational softening length of $700$ proper pc \citep[see also][]{bs12}.
The high resolution of our zoom-ins allows us to resolve the main structures in the ISM of galaxies. 
Our simulations also explicitly treat stellar feedback from supernovae of Types I and II, stellar winds from young and evolved stars, photoionization, and radiation pressure on dust grains. 

Our suite of simulations has been shown to successfully reproduce the observationally-inferred relationship between stellar mass and dark matter halo mass \citep[the $M_{\star}-M_{\rm h}$ relation][]{2014MNRAS.445..581H} and the mass-metallicity relations \citep[][]{2016MNRAS.456.2140M} of galaxies below $\sim L^{\star}$ at all redshifts where observational constraints are currently available, as well as the covering fractions of dense HI in the halos of $z=2-4$ Lyman break galaxies (LBGs) and quasars \citep[][]{2015MNRAS.449..987F, 2016arXiv160107188F}. 
In these simulations, we find that star formation histories and the resultant galactic winds are highly time variable (Muratov et al. 2015\nocite{2015MNRAS.454.2691M}; Sparre et al. 2015\nocite{2015arXiv151003869S}), a property that we show in this paper induces corresponding time variability in the UV emission from circum-galactic gas. 
Unlike in some more \emph{ad hoc} stellar feedback implementations, our simulations follow hydrodynamical interactions and gas cooling at all times, lending some credence to the phase structure of galactic winds predicted in our calculations. 

Our primary goal in this paper is to investigate the implications of the high resolution and explicit stellar feedback models of the FIRE simulations for UV metal line emission from the CGM of high-redshift galaxies and its physical origin. 
Our analysis is guided by the UV metal lines most likely to be detectable by MUSE and KCWI, though it is beyond the scope of this paper to make detailed observational predictions for specific instruments.

This paper is organized as follows. 
We describe our simulations in more detail and our methodology for computing metal line emission in \S \ref{methodology}. 
We present our main results 
in \S \ref{results}, including a discussion of the strong predicted stellar feedback-driven UV metal line time variability. 
We summarize our conclusions in \S \ref{s:conclusions}. 
The Appendices contain a convergence test and local source test. 
\section{Methodology}
\label{methodology}

\subsection{Simulations} \label{s:simulations}

\begin{figure*}
\begin{center}
\includegraphics[width=\textwidth]{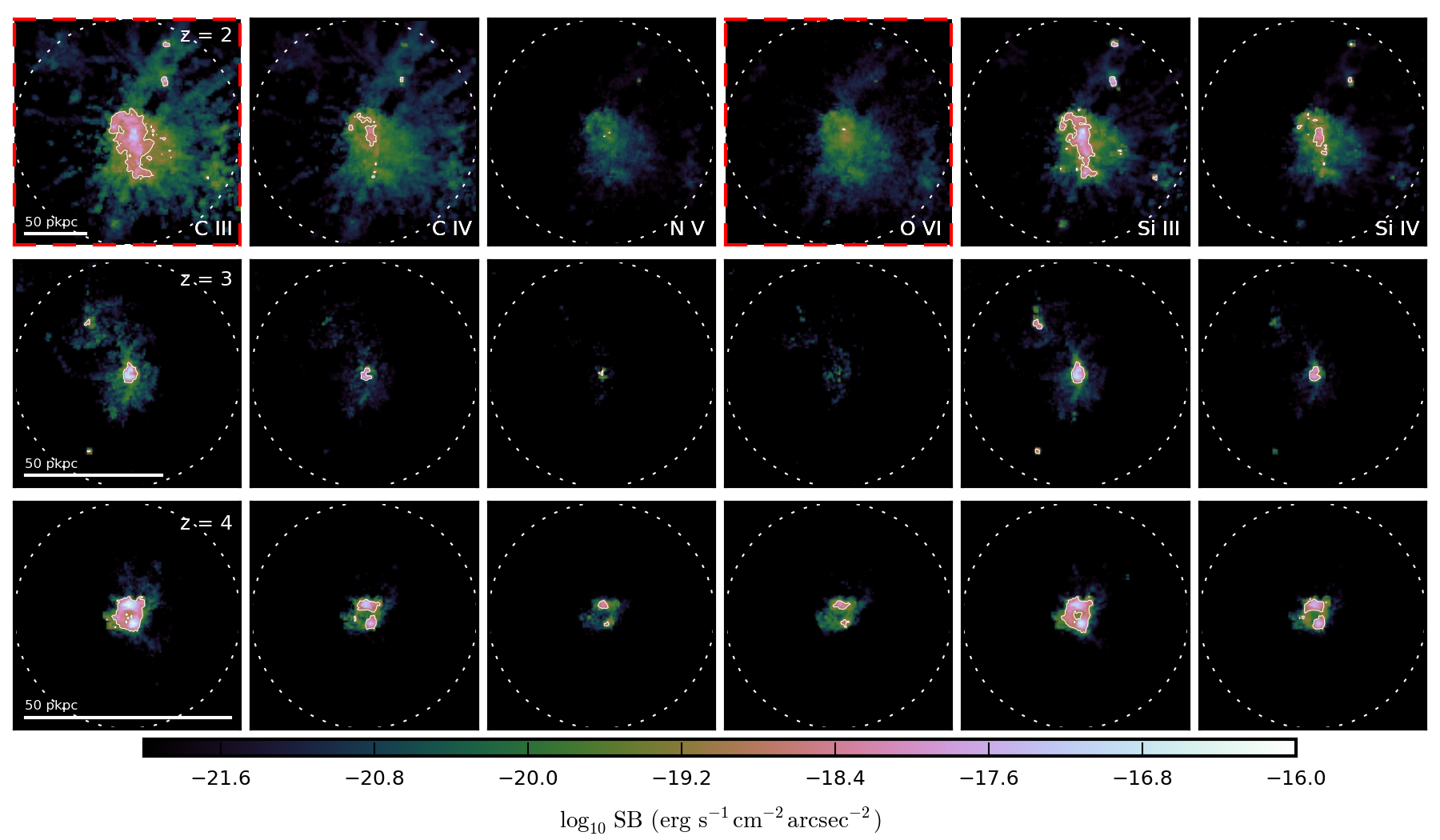}
\end{center}
\caption[]{
Theoretical surface brightness maps for the UV metal lines studied in this work in {\bf z2h506} in a region with side length equal to a virial radius at $z=2$ (top), 3 (middle), and 4 (bottom). 
The halo mass in $\log_{10}M_{\rm h}$ (M$_\odot$) is 12.06  ($z = 2$), 11.36 ($z = 3$), and 11.11 ($z = 4$). 
From left to right, the vertical panels at each redshift show the emission for the \ciii, \civ, \nv, \ovi, \siiii, and \siiv~lines listed in Table \ref{t:lines}. 
The dashed lines show the virial radius in each panel. The solid white contours enclose regions with SB $> 10^{-19}$~erg s$^{-1}$~cm$^{-2}$~arcsec$^{-2}$, a proxy for the SB detectable in (deep but non-stacked) MUSE and KCWI observations. 
All the emission lines shown would redshift into either the MUSE or KCWI spectral ranges, except  C III and O VI at $z=2$ (indicated by dashed red borders around the panels). 
The observed surface brightness for the CIII, SiIII, and OVI lines is subject to IGM attenuation (10-20\%, 30-50\%, and factor 2-5 effects at $z=2,~3,~{\rm and}~4$, respectively; \S \ref{sec:igm_absorption}).
}
\label{f:h506_em_maps_19} 
\end{figure*}

\begin{figure*}
\begin{center}
\includegraphics[width=\textwidth]{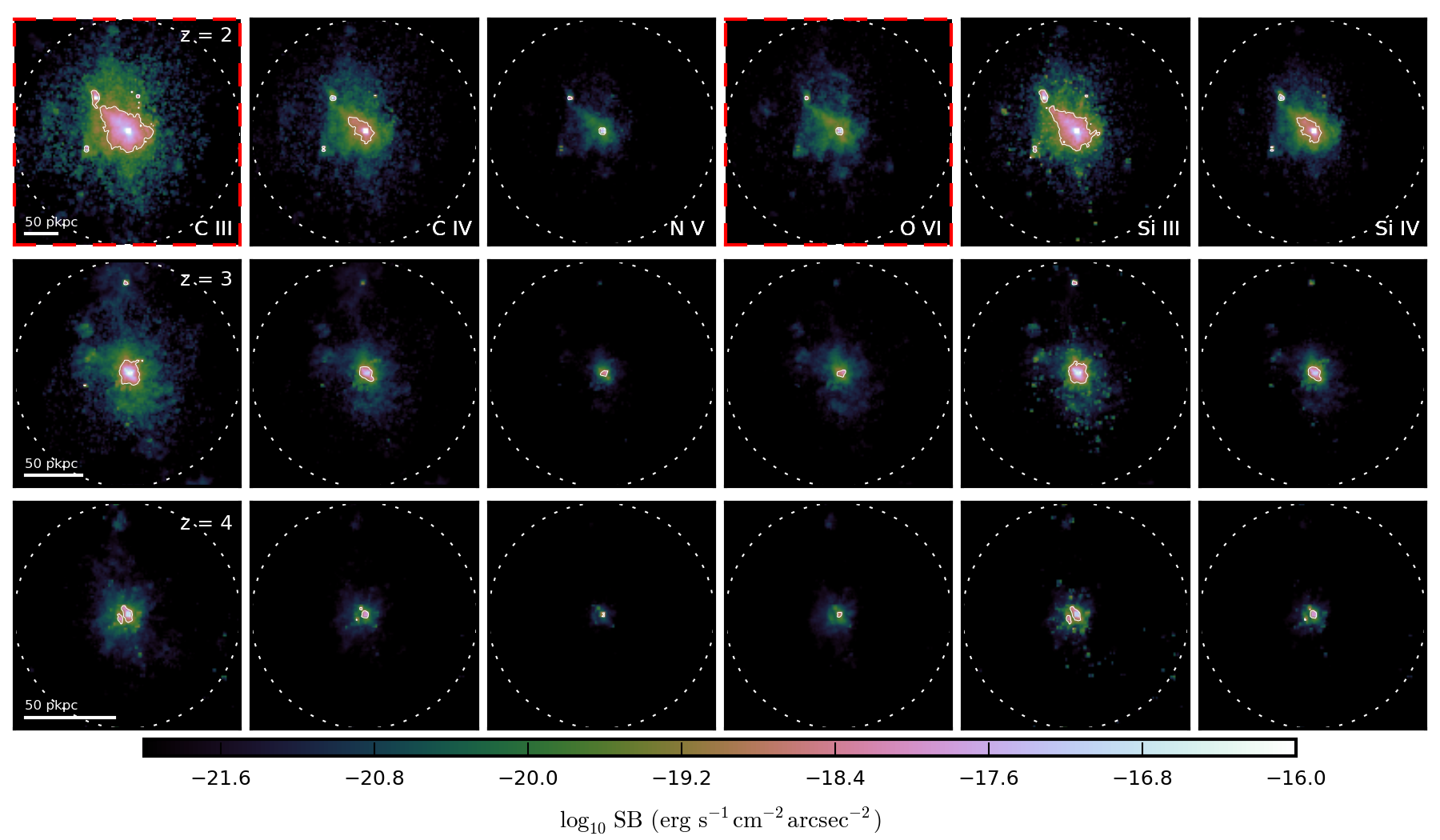}
\end{center}
\caption[]{
Same as in Figure \ref{f:h506_em_maps_19} but for the more massive halo {\bf MFz2\_B1}. 
The halo mass in $\log_{10}M_{\rm h}$ (M$_\odot$) is 12.93  ($z = 2$), 12.52 ($z = 3$), and 12.21 ($z = 4$).
}
\label{f:m14_em_maps_19} 
\end{figure*}

Here we summarize the numerical methods used in our simulations and the physics included. 
For more details on the algorithms, we refer the reader to \citet{2014MNRAS.445..581H} and references therein.

Our simulations were run with the GIZMO code \citep[][]{2014arXiv1409.7395H} in ``P-SPH'' mode. 
P-SPH, described in \cite{2013MNRAS.428.2840H}, is a Lagrangian-based pressure-entropy formulation of the smooth particle hydrodynamics (SPH) equations that eliminates some of the well-known differences between traditional implementations of SPH and grid-based codes \citep[e.g.,][]{2007MNRAS.380..963A, 2012MNRAS.424.2999S} while preserving the excellent conservation properties of SPH. 
GIZMO derives from Gadget-3 \citep[last described in][]{2005MNRAS.364.1105S} and its tree particle-mesh (TreePM) gravity solver is based on the latter. 
Our GIZMO runs include improved algorithms and prescriptions for artificial viscosity \citep{2010MNRAS.408..669C}, entropy mixing \citep{2008JCoPh.22710040P}, adaptive time stepping \citep{2012MNRAS.419..465D}, the smoothing kernel \citep{2012MNRAS.425.1068D}, and adaptive gravitational softening \citep{pm07, b12}. 
Star formation in the FIRE simulations only occurs in molecular, self-gravitating gas with $n_{\rm H} \gtrsim 5 - 50$ cm$^{-3}$. The energy, momentum, mass, and metal yields from photo-ionization, radiation, stellar winds and supernovae are calculated using STARBURST99 \citep{1999ApJS..123....3L}. 
The abundances of nine metallic species (C, N, O, Ne, Mg, Si, S, Ca, and Fe) are also tracked using this method and used to evaluate cooling rates using a method similar to \cite{2009MNRAS.393...99W}. Ionization balance of these elements are computed assuming the cosmic ionizing background of \citet{2009ApJ...703.1416F} (FG09).\footnote{Publicly available at http://galaxies.northwestern.edu/uvb.} Self-shielding of hydrogen is accounted for with a local Jeans-length approximation (integrating the local density at a given particle out to a Jeans length to determine a surface density $\Sigma$), then attenuating the background seen at that point by $e^{-\kappa \Sigma}$ (where $\kappa$ is the opacity). Confirmation of the accuracy of this approximation in radiative transfer experiments can be found in \cite{fkd10}.

The simulations analyzed in this work are listed in Table \ref{t:simulations}. 
The initial conditions for {\bf m13}, {\bf m12q}, and {\bf m12i} were chosen to match ones from the AGORA project \citep{2014ApJS..210...14K}. 
The initial conditions for {\bf m12v} are the same as the `B1' run studied by \cite{2009ApJ...700L...1K} and \citet{2011MNRAS.412L.118F}. 
The {\bf z2hxxx} series of simulations, first described in \citet{2015MNRAS.449..987F}, consists of halos in the mass range $M_{\rm h} = 1.9\times10^{11}$ M$_\odot~-~1.2\times10^{12}$ M$_\odot$ at $z=2$. 
These halos are representative of ones hosting LBGs at that redshift, though somewhat on the low-mass end of the mass distribution  \citep[LBGs at $z\sim2$ reside in dark matter halos of average mass $M_{\rm h}\sim10^{12}$ M$_{\odot}$;][]{2005ApJ...619..697A, 2012ApJ...752...39T}. 
{\bf MFz2\_B1} is part of the MassiveFIRE simulation suite (Feldmann et al. in prep). 
Our simulations do not include feedback from AGN. While such feedback may be potentially important, especially in the more massive halos, the proper modeling of AGN feedback is still a major theoretical challenge \citep[e.g.,][]{2014arXiv1412.2712S} and we have decided to postpone its treatment to future work. 
We focus our analysis on main halos (i.e., we do not center on satellite halos). 
The surface brightness profiles of UV line emission may, however, include emission contributed by satellite galaxies. 
Halos are identified using Amiga's Halo Finder \citep[AHF;][]{2009ApJS..182..608K} and we adopt the virial overdensity definition of \cite{1998ApJ...495...80B}. 

All our simulations assume a ``standard'' flat $\Lambda$CDM cosmology with $h\approx0.7$, $\Omega_{\rm m} = 1-\Omega_{\Lambda} \approx 0.27$ and $\Omega_{\rm b}\approx0.046$. Minor variations about these fiducial values are adopted for some simulations to match the parameters of simulations from the AGORA project and from our previous work. Uncertainties in our calculations are dominated by baryonic physics and the small variations in cosmological parameters do not introduce significant effects in our analysis. 

\subsection{Emissivity Calculation} \label{s:embs12_tables}
Our method for calculating gas emissivities is similar to the one used by \citet{vs13}. 

\begin{footnotesize}
\ctable[caption={{\normalsize UV Lines Analyzed in this Work}}\label{t:lines},center,nostar]
{lcccccc}
{\tnote[ ]{
{\bf (1)} $\lambda_1$: Rest-frame wavelength of the emission line or the stronger line for doublets. \\
{\bf (2)} $\lambda_2$: Rest-frame wavelength of the weaker line for doublets. \\
{\bf (3)} $z_{\rm min}$: The minimum redshift from which the stronger transition line lies within the wavelength coverage of MUSE and KCWI.\\
{\bf (4)} $z_{\rm max}$: The maximum redshift from which the stronger transition line lies within the wavelength coverage of MUSE and KCWI. \\
Note: The wavelength coverage of MUSE and KCWI are 4650\AA\ -- 9300\AA\ and 3500\AA\ -- 10,500\AA, respectively.
}}
{
\hline\hline
&&& \multicolumn{2}{c}{MUSE} & \multicolumn{2}{c}{KCWI}\\
\multicolumn{1}{l}{Ion Name} &
\multicolumn{1}{c}{$\lambda_1$} &
\multicolumn{1}{c}{$\lambda_2$} & 
\multicolumn{1}{c}{$z_{\rm min}$} & 
\multicolumn{1}{c}{$z_{\rm max}$} & 
\multicolumn{1}{c}{$z_{\rm min}$} & 
\multicolumn{1}{c}{$z_{\rm max}$} \\ 
\multicolumn{1}{l}{\ } &
\multicolumn{1}{c}{\AA} &
\multicolumn{1}{c}{\AA} &
\multicolumn{1}{c}{\ } &
\multicolumn{1}{c}{\ } &
\multicolumn{1}{c}{\ } &
\multicolumn{1}{c}{\ } \\ 
\hline
\civ		& 1548 & 1551 & 2.00 & 5.01   & 1.26 & 5.78 \\
\siiv		& 1394 & 1403 & 2.34 & 5.67   & 1.51 & 6.53 \\
\nv		& 1239 & 1243 & 2.75 & 6.51   & 1.82 & 7.47 \\
\ovi		& 1032 & 1038 & 3.51 & 8.01   & 2.39 & 9.17 \\
\siiii		& 1207 & -        & 2.85 & 6.71   & 1.90 & 7.70 \\
\ciii		& 977   & -        & 3.76 & 8.52   & 2.58 & 9.75 \\
\hline\hline\\
}
\end{footnotesize}

Following previous work that identified the most important contributions to rest-UV emission from the high-redshift CGM \citep[][]{bs12, vs13}, we compute emissivities for UV doublets \civ\ (1548 \AA, 1551 \AA), \siiv\ (1394 \AA, 1403 \AA), \nv\ (1239 \AA, 1243 \AA), and \ovi\ (1032 \AA, 1038 \AA), and the singlets \siiii\ (1207 \AA) and \ciii\ (977 \AA). 
{We do not show predictions for the \neviii\ (770 \AA, 780 \AA) doublet because its intrinsic CMG luminosity is $\sim$3-4 dex below the CIII line and suffers from Lyman continuum absorption by the IGM. We do not expect that this line will be observable in the near future. 
For each doublet, we report only predictions for the strongest line in the doublet. 
For these doublets, the ratio of the intensity of the stronger and weaker lines in the doublet is $\approx2$. 
Therefore, the total intensity for the doublet (e.g., as would be measured by an instrument that does not spectrally resolve the two transitions) can be obtained by multiplying the intensities predicted by a factor of $\approx1.5$.  
Table \ref{t:lines} lists the emission lines we study and the minimum and maximum redshifts at which they lie within the wavelength coverage of MUSE and KCWI. 
We first pre-compute grids of line emissivities as a function of gas temperature and hydrogen number density over the redshift interval $z = 2 - 4$ using \cloudy\ (version 13.03 of the code last described by Ferland et al. 2013\nocite{fpv13}). 
The grids sample temperatures in the range $10< T < 10^{9.5}~{\rm K}$ in intervals of $\Delta \log_{10}{T} = 0.05$ and hydrogen number densities in the range $10^{-8} < n_{\rm H} < 10^4~{\rm cm^{-3}}$ in intervals of $\Delta \log_{10}{n_{\rm H}}$ = 0.2. 
The grids are computed at redshift intervals of $\Delta z = 0.1$. 
For the grids, the gas is assumed to be of solar metallicity, optically thin and in photoionisation equilibrium with the cosmic ionizing background (the contribution of collisional ionization to equilibrium balance is automatically included). 
In computing emissivities from our simulations, we bi-linearly interpolate the pre-computed grids in logarithmic space and scale linearly with the actual metallicity of the gas. 
For most of our calculations, we use the cosmic ionizing background model of \cite{2009ApJ...703.1416F}; we show in \S \ref{s:SB_prof} that the choice of ionizing background model and photoionization from local sources affect our predictions only weakly, indicating that most of the UV emission from metal lines in the CGM originates from regions that are collisionally ionized \citep[see also][]{vs13}. 

Non-equilibrium ionization and cooling effects \citep[e.g.,][]{2007ApJS..168..213G, 2013MNRAS.434.1043O} are not captured in the present calculations.

\subsection{Surface Brightness Calculation} \label{s:em_calc}
In order to make full use of the spatially-adaptive resolution of our Lagrangian SPH simulations, we evaluate line luminosities from the particle data directly. 
For a given gas particle, the luminosity of each line listed in Table \ref{t:lines} is calculated as
\begin{equation} 
\label{eq:Lpart}
L_{\rm part} = \epsilon_\odot(z, n_{\rm H}, T)~\left( \frac{m_{\rm gas}}{\rho_{\rm gas}} \right)~\left( \frac{Z_{\rm gas}}{Z_\odot} \right),
\end{equation}
where $m_{\rm gas}$, $\rho_{\rm gas}$, and $Z_{\rm gas}$ are the mass, density and metallicity of the gas particle, respectively, and $\epsilon_\odot(z, n_{\rm H}, T)$ is the emissivity interpolated from the pre-computed solar-metallicity grid. 
This approach for evaluating emissivities is based on total gas metallicity and assumes that the line emitting gas has solar abundance ratios. 
Our SPH simulations do not include a model for metal diffusion by unresolved turbulence \citep[e.g.,][]{2010MNRAS.407.1581S}. While this could potentially cause the metals in the simulations to be too clumped and artificially enhance our predicted luminosities, our simulations do capture the mixing of metals due to resolved gas flows. 
In Appendix \ref{sec:convergence}, we present a convergence test showing that our predicted luminosities do not increase significantly with increasing resolution (and hence increased metal mixing).

\begin{figure*}
\begin{center}
\includegraphics[width=\textwidth]{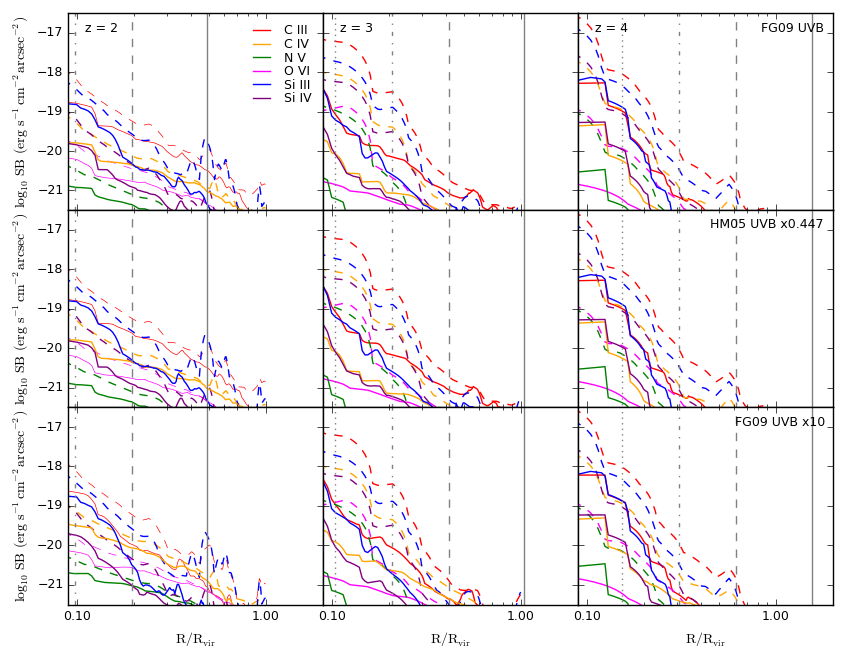}
\end{center}
\caption[]{Median (solid lines) and mean (dotted lines) UV metal line theoretical radial surface brightness profiles for the sample of Lyman break galaxy simulations described in \S~\ref{s:simulations} (excluding {\bf MFz2\_B1}) at $z = 2$ (left), 3 (middle), and 4 (right), assuming the FG09 UV/X-ray ionizing background model (top), the galaxy+quasar HM05 UV/X-ray background scaled by a factor of 0.447 (to match the hydrogen photoionization rate of the FG09 model at $z=3$; middle), and the FG09 background multiplied by a factor of 10 (bottom). 
The mean and median halo masses in $\log_{10}M_{\rm h}$ (M$_\odot$) are 11.79 and 11.82 ($z = 2$),  11.44 and 11.40 ($z = 3$), and 11.11 and 11.10 ($z = 4$), respectively. 
The vertical dotted, dot-dashed, dashed and solid grey lines indicate radii of 5, 10, 20, and 50 proper kpc for median-mass halos at each redshift. 
Surface brightness profiles shown with thinner curves correspond to emission lines that do not redshift into the MUSE or KCWI bands (indicated by the dashed red border panels in Figs. \ref{f:h506_em_maps_19} and \ref{f:m14_em_maps_19}). The observed surface brightness for the CIII, SiIII, and OVI lines is subject to IGM attenuation (10-20\%, 30-50\%, and factor 2-5 effects at $z=2,~3,~{\rm and}~4$, respectively; \S \ref{sec:igm_absorption}).
}
\label{SB_all} 
\end{figure*}

To produce surface brightness profiles, we assign particle luminosities to 3D Cartesian grids. 
The flux from each grid cell 
is given by
\begin{equation} 
F_{\rm cell} = \frac{L_{\rm cell}}{4 \pi d_{\rm L}^2},
\end{equation}
where $d_{\rm L}$ is the luminosity distance and $L_{\rm cell} = \sum L_{\rm part}$ is the sum of the particle luminosities assigned to the cell. 
The cell side length of the grids is 1 proper kpc for all simulations. 
{This introduces smoothing on that spatial scale (corresponding to 0.1 arcsec at $z=2$) but does not bias the surface brightness calculations since luminosities are computed from the un-degraded particle-carried information. 
Finally, the surface brightness (SB) projected on the sky 
is calculated by dividing the flux from each grid cell by the solid angle $\Omega_{\rm cell}$ it subtends and summing along the line of sight (los):
\begin{equation} 
SB = \sum\limits_{\rm los}\frac{F_{\rm cell}}{\Omega_{\rm cell}}.
\end{equation}
In the small angle limit and for cubical grid cells, $\Omega_{\rm cell} = (a_{\rm cell} / d_{\rm A})^{2}$, 
where 
$a_{\rm cell}$ is the proper side length of a grid cell and $d_{\rm A}$ is the angular diameter distance.

\subsection{IGM Attenuation}
\label{sec:igm_absorption}
Absorption by intervening intergalactic gas affects the detectability of the emission lines we predict \citep[e.g.,][]{1995ApJ...441...18M}. 
The emission lines we analyze in this paper all have rest wavelength $\lambda>912$~\AA~and so are not affected by Lyman continuum opacity. The CIII, SiIII, and OVI emission lines that we consider however have rest wavelengths $\lambda<1216$~\AA~and so will be attenuated by Lyman-series opacity. 
Using the IGM attenuation model of \cite{2014MNRAS.442.1805I}, we calculated the transmission factor $e^{-\tau}$ for the CIII 977~\AA, SiIII 1207~\AA, and OVI 1032~\AA~lines. 
For these three lines, we find transmission factors $(0.9,~0.8,~0.9)$ at $z=2$, $(0.7,~0.5,~0.7)$ at $z=3$, and $(0.5,~0.2,~0.4)$ at $z=4$, respectively. 
Thus, IGM attenuation has a small effect on our predictions at $z=2$. 
Even out to $z=4$, IGM attenuation is only a factor $\sim2$ effect for CIII and OVI. 
While significant, this is much smaller than the order-of-magnitude time variability that our simulations predict (see \S \ref{s:time_var}), so this does not significantly affect any of our main conclusions. 
At $z=4$, the IGM may suppress observable SiIII emission by a factor of $\sim5$, which will make this line more challenging to detect at such redshifts. 
The CIV, NV, and SIV lines we analyze do not suffer from Lyman-series absorption and so are not significantly affected by hydrogen absorption from the IGM. 
To allow readers to easily correct our results for other IGM attenuation models (proximity effects could be important for SiIII 1207~\AA, which has a rest wavelength close to Ly$\alpha$ 1216~\AA), in this paper we plot theoretical surface brightnesses taking into account cosmological dimming but neglecting IGM attenuation.

\section{Results}
\label{results}

\begin{figure*}
\begin{center}
\includegraphics[width=\textwidth]{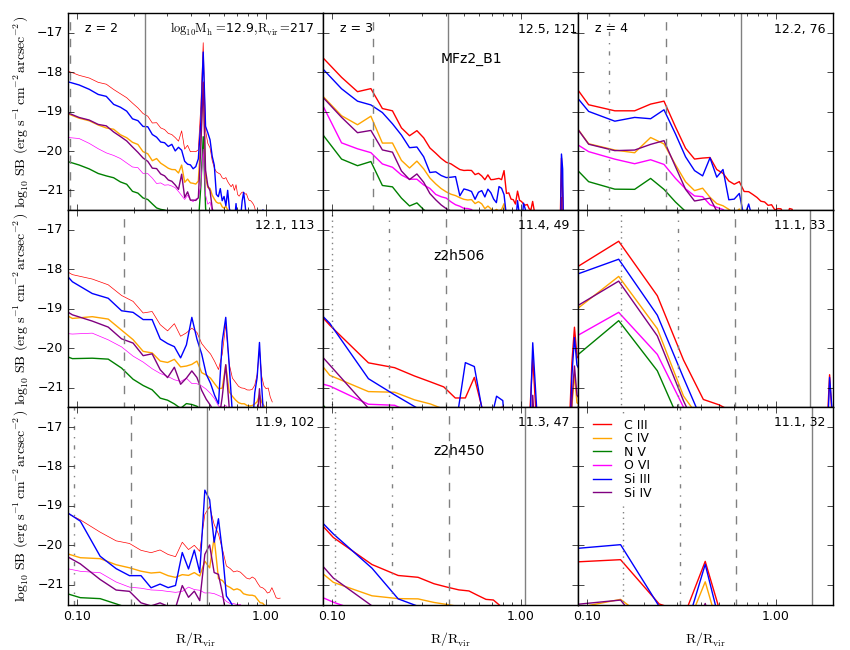}
\end{center}
\caption[]{Cylindrically averaged UV metal line radial theoretical surface brightness profiles for halos {\bf MFz2\_B1} (top), {\bf z2h506} (middle), and {\bf z2h450} (bottom) at $z = 2$ (left), 3 (middle), and 4 (right). Values of $\log_{10}~M_{\rm h}(M_\odot)$ and $R_{\rm vir}$ (proper kpc) for each halo are indicated at the top right of each panel. The vertical dotted, dot-dashed, dashed and solid grey lines indicate radii of 5, 10, 20, and 50 proper kpc, respectively. Surface brightness profiles shown with thinner curves correspond to emission lines that do not redshift into the MUSE or KCWI bands (indicated by the dashed red border panels in Figs. \ref{f:h506_em_maps_19} and \ref{f:m14_em_maps_19}). The observed surface brightness for the CIII, SiIII, and OVI lines is subject to IGM attenuation (10-20\%, 30-50\%, and factor 2-5 effects at $z=2,~3,~{\rm and}~4$, respectively; \S \ref{sec:igm_absorption}).}
\label{rad_prof} 
\end{figure*}

\subsection{UV Metal Lines Maps and Radial Profiles} \label{s:SB_prof}
Figures \ref{f:h506_em_maps_19} and \ref{f:m14_em_maps_19} show surface brightness maps for the UV metal lines in Table \ref{t:lines} for {\bf z2h506} and {\bf MFz2\_B1}, respectively. 
These are our two most massive halos at $z=2$. 
In these maps, solid white contours enclose regions with SB $> 10^{-19}$~erg s$^{-1}$~cm$^{-2}$~arcsec$^{-2}$, a proxy for the SB detectable in (deep but non-stacked) MUSE and KCWI observations. 
High-redshift \lya\ CGM emission has been detected in narrow-band imaging observations 
 down to $\sim10^{-18}$~erg s$^{-1}$ cm$^{-2}$ arcsec$^{-2}$ in individual objects \citep{2000ApJ...532..170S, 2004AJ....128..569M} and down to $\sim10^{-19}$~erg s$^{-1}$ cm$^{-2}$ arcsec$^{-2}$ in stacks \citep{2011ApJ...736..160S}. 
Current and upcoming IFSs such as MUSE and KCWI are expected to reach an order of magnitude deeper. 
For reference, SB $= 10^{-19}$~erg s$^{-1}$~cm$^{-2}$~arcsec$^{-2}$ is the $1\sigma$ detection threshold for the azimuthally averaged Ly$\alpha$ radial profiles in the recent ultra-deep exposure of the Hubble Deep Field South obtained with MUSE \citep[][]{2015arXiv150905143W}. 
Comparing Figures \ref{f:h506_em_maps_19} ($M_{\rm h}=1.2\times10^{12}$ M$_{\odot}$ at $z=2$) and \ref{f:m14_em_maps_19} ($M_{\rm h}=8.5\times10^{12}$ M$_{\odot}$ at $z=2$) suggests a significant halo mass dependence for the detectability of metal UV lines, with the more massive halo showing more spatially extended and more luminous emission (note the different spatial scales). 
In \S \ref{s:time_var} we will show that the CGM UV metal line emission is primarily powered by energy injection from galactic winds. 
Since more massive halos on average have higher star formation rates (before quenching), these halos are naturally more luminous \emph{on average}. 

For {\bf MFz2\_B1}, the C III and Si III lines are predicted to produce emission above $10^{-19}$~erg s$^{-1}$ cm$^{-2}$ arcsec$^{-2}$ on spatial scales $\sim100$ proper kpc (in the elongated direction) at $z=2$, and C IV and Si IV should be detectable at this surface brightness on scales $\sim 50$ proper kpc. 
For {\bf z2h506}, we predict similar structures but with spatial extent smaller by approximately the ratio of the virial radius in {\bf z2h506} relative to that of {\bf MFz2\_B1} (a factor of $\approx0.5$ at $z=2$). 
These transitions preferentially probe relatively cool, $T\sim10^{4.5}-10^{5}$ K gas.
All these lines would redshift into either the MUSE or KCWI spectral bands if observed from $z=2-4$, except C III at $z\sim2$. 
In Figure \ref{SB_all}, we show the median and mean radial surface brightness profiles\footnote{To calculate the average mean and median radial profiles, we first compute the cylindrically averaged profiles for individual halos, then take the mean and median over halos in our sample. To compute the cylindrical averages, the radial profiles are radially binned in bins of width 3 proper kpc.} for all the halos in our sample but excluding excluding {\bf MFz2\_B1} (which is of substantially higher mass than the rest). 
O VI (which preferentially probes warmer gas, $T\sim10^{5.5}$ K gas) is less luminous than C IV and Si IV by $\sim1$ dex and does not redshift into the KCWI and MUSE bands at $z=2$. 
O VI emission may be detectable at higher redshifts in very deep integrations of massive halos but could prove beyond the reach of the current generation of instruments. 
As we show in \S \ref{s:time_var}, UV emission from the CGM is highly time variable. 
Therefore, different halos of the same mass and at the same redshift may produce very different circum-galactic UV metal emission profiles.

The metal UV line emission from our simulated halos is generally easier to detect at $z\sim2$ than at $z\sim3-4$. 
Two effects contribute to this. 
First, our halos grow with time and their star formation rates also on average increase with time over that redshift interval \citep[e.g.,][]{2014MNRAS.445..581H, 2015MNRAS.454.2691M}, thus increasing the amount of energy injected in the CGM and radiated by the UV lines. 
Second, the surface brightness is subject to cosmological surface brightness dimming, $SB \propto (1+z)^{-4}$. 
Note that the virial radii of the simulated halos in Figure \ref{SB_all} increase with decreasing redshift so that the total luminosities of the $z=2$ halos are systematically higher even though the surface brightness at a fixed fraction of $R_{\rm vir}$ appears to follow a different redshift trend.

The different rows in Figure \ref{SB_all} correspond to different assumptions for the ionizing flux illuminating CGM gas. 
The top row assumes our fiducial FG09 UV/X-ray background.  
To test the sensitivity to the shape of the ionizing spectrum, we have repeated our CLOUDY emissivity calculations using the Haardt \& Madau 2005 background, which we rescaled by a factor of 0.447 to match the hydrogen photoionization rate $\Gamma_{\rm HI}=5.5\times10^{-13}$ s$^{-1}$ of the FG09 model at $z=3$ (middle row). 
We also tested the sensitivity of our results to the magnitude of the ionizing flux by multiplying the FG09 background by a factor of 10 (bottom row). 
The mean and average surface brightness profiles are almost identical between the different rows, at $z=2,~3$ and 4, indicating that our predictions are not sensitive to the assumed ionizing background model. 
This implies that the UV metal line emission arises primarily from collisionally ionized gas whose ionization state is not sensitive to the ionizing flux, as previously found by \cite{vs13}. 
Furthermore, the metal line emission that we focus on arises primarily from gas at temperatures $T\sim10^{4.5}-10^{6}$ K, which is generally collisionally heated. 
We have verified that our simulations predict emissivity-weighted temperatures broadly consistent with Figure 6 of \cite{vs13}. 
This implies that self-shielding effects do not significantly affect our UV metal line emission, unlike for Ly$\alpha$ emission for which self-shielding effects are critical \citep[e.g.,][]{fkd10}. This experiment also indicates that ionizing photons emitted by local galaxies (which we have neglected) likely would not substantially alter our predictions. 
Measurements of the fraction of ionizing photons that escape the ISM of LBGs at $z\sim3$ moreover indicate that $f_{\rm esc}\sim5\%$ \citep[e.g.,][]{2006ApJ...651..688S}. 
If this fraction corresponds to the fraction of directions along which ionizing photons escape relatively unimpeded, and the ISM is opaque to ionizing photons along other directions, most of the CGM around star-forming galaxies is not strongly illuminated by the central galaxy. 
In Appendix \ref{sec:local_xrays}, we present another test explicitly including a local source spectrum (including X-rays from supernova remnants) that also shows that our main predictions are not significantly affected by local sources (though some details of the line emission at low surface brightnesses are).

\begin{figure*}
\begin{center}
\includegraphics[width=\textwidth, trim = 0.3in 0.1in 0.1in 0.4in, clip=true]{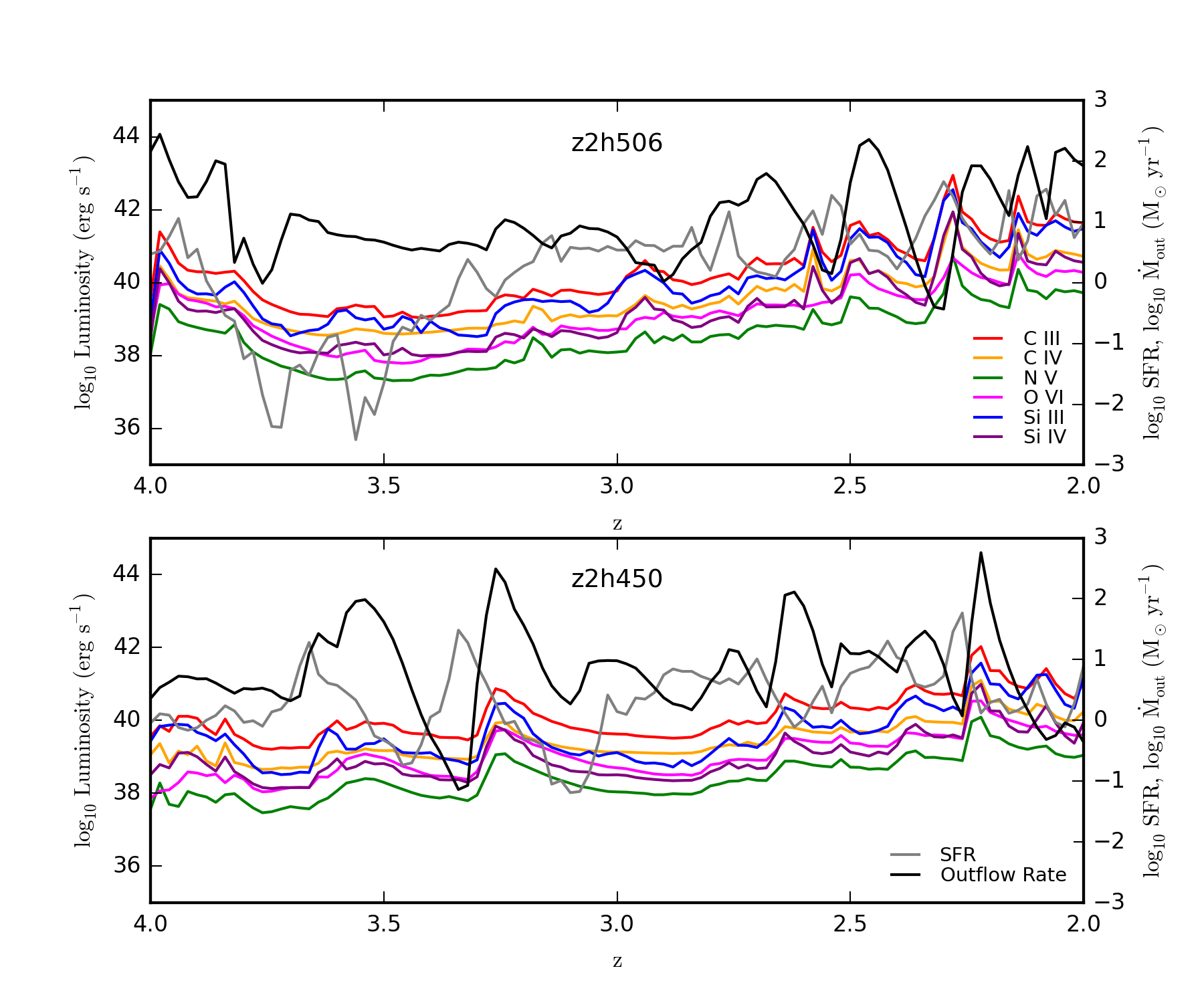}
\end{center}
\caption[]{Colored lines show UV metal line luminosities within $1~R_{\rm vir}$ but excluding the inner 10 proper kpc 
(a proxy for central galaxies) in simulated halos {\bf z2h506} (top) and {\bf z2h450} (bottom) as a function of redshift. 
Star formation rates within $1~R_{\rm vir}$ and gas mass outflow rates at $0.25~R_{\rm vir}$ as a function of redshift are plotted as grey and black lines, respectively. 
The UV metal line luminosities, star formation, and mass outflow rates are all strongly time variable and correlated. 
Peaks in CGM luminosity correspond more closely with peaks in mass outflow rates, which follow peaks of star formation with a time delay, indicating that energy injected in the halos by galactic winds powers the UV metal line emission. 
}
\label{time_evol} 
\end{figure*}

To give a sense of how our predictions vary from halo to halo, Figure \ref{rad_prof} shows the predicted cylindrically averaged radial surface brightness profiles for a representative sample of individual halos ({\bf MFz2\_B1}, {\bf z2h506}, and {\bf z2h450}). 
The spikes in the radial profiles (for example, at $R/R_{\rm vir}\sim$~0.5 for {\bf MFz2\_B1} at $z = 2$) are due to satellite galaxies (see the maps in Figs. \ref{f:h506_em_maps_19} and \ref{f:m14_em_maps_19}). 

Our predicted average SB profiles shown in the top panels of Figure \ref{SB_all} agree with those obtained by \citet{vs13} from the OWLS simulations to within $\sim$ 0.5 dex on average (with no apparent systematic differences). 
However, we find that the intensities of UV metal lines vary by several dex from halo to halo. Furthermore, there is strong variability in the strength of individual metal lines between halos of otherwise similar mass and redshift. 
For example, the strength of \ciii\ in {\bf z2h506} and {\bf z2h450} at $z = 4$ differ by $\sim$ 3 dex in surface brightness, even though both halos have a mass $M_{\rm h}\approx10^{11.1}$ M$_{\odot}$ at that redshift. 
We address this in the next section.

\begin{figure*}
\begin{center}
\includegraphics[width=\textwidth, trim = 0.3in 0.1in 0.1in 0.4in, clip=true]{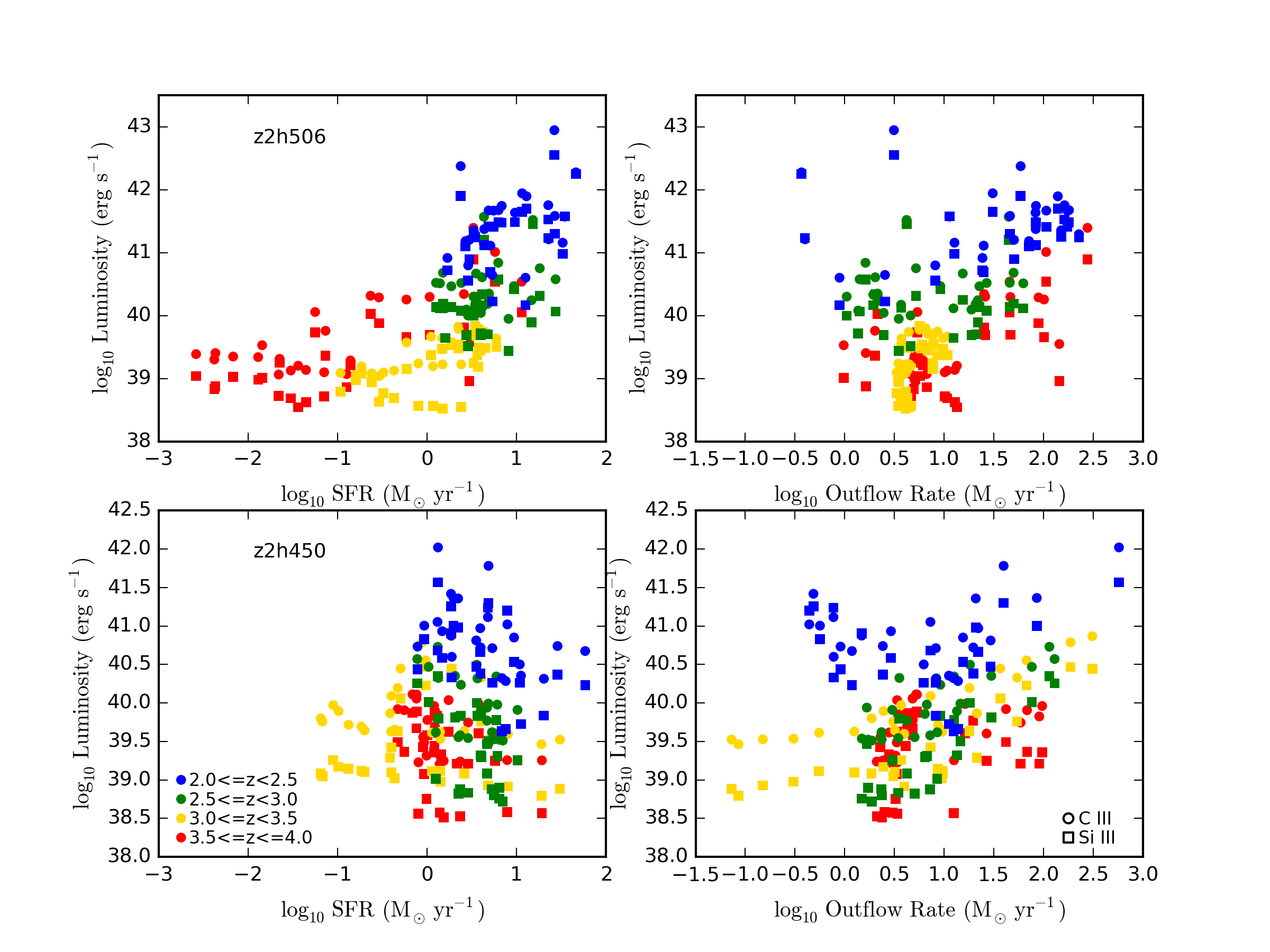}
\end{center}
\caption[]{UV metal line luminosities within $1~R_{\rm vir}$ but excluding the inner 10 proper kpc
of halo centers (a proxy for central galaxies) in simulated halos {\bf z2h506} (top) and {\bf z2h450} (bottom). 
The luminosities are plotted versus the star formation rate within $R_{\rm vir}$ (left) and versus the mass outflow rate at $0.25 R_{\rm vir}$. 
Circle symbols show the CIII luminosity and square symbols show SiIII luminosity. 
Colors indicate redshift bins (100 time slices between $z=4$ and $z=2$ are shown). 
Overall, CGM UV line luminosities correlate positively with mass outflow rate at fixed redshift but no significant correlation with instantaneous star formation rate is apparent.
}
\label{fig:correlations} 
\end{figure*}

\subsection{Star Formation-Driven Time Variability} \label{s:time_var}
One of the key predictions of the FIRE simulations with resolved ISM is that the star formation histories of galaxies have a strong stochastic component and are much more time variable (e.g., Hopkins et al. 2014; Sparre et al. 2015\nocite{2015arXiv151003869S}) than in lower-resolution simulations in which the ISM is modeled with a sub-grid effective equation of state. 
Such sub-grid equations of state are standard in current large-volume cosmological simulations \citep[e.g.,][]{2013MNRAS.434.2645D, 2014MNRAS.444.1518V, 2015MNRAS.446..521S}. 
In particular, a sub-grid equation of state was used in the OWLS simulations analyzed by \cite{vs13} for UV line emission from the CGM. 
We have previously shown that the time variability of star formation in the FIRE simulations results in time variable galactic winds \citep[][]{2015MNRAS.454.2691M}, an increased cool gas content of galaxy halos \citep[][]{2015MNRAS.449..987F}, and can transform dark matter halo cusps into cores in dwarf galaxies (Onorbe et al. 2015\nocite{2015arXiv150202036O}; Chan et al. 2015\nocite{2015arXiv150702282C}). 
Similar time variability has also been found in other zoom-in simulations implementing different stellar feedback models \citep[][]{2010Natur.463..203G, 2011ApJ...742...76G, 2013MNRAS.428..129S, 2014arXiv1404.2613A}, indicating that it is a generic consequence of resolving the ISM of galaxies, in which local dynamical time scales can be very short. 
We now show that time-variable star formation also has important implications for the observability of CGM gas in emission, and explains why two halos of the same mass and redshift can vary in metal UV line luminosity by orders of magnitude.

In Figure \ref{time_evol} we plot the luminosities of UV metal lines within $1R_{\rm vir}$ 
(a proxy for emission from central galaxies), for {\bf z2h506} and {\bf z2h450}, as a function of redshift. 
We also plot the star formation rates within $1R_{\rm vir}$ and gas mass outflow rates at $0.25R_{\rm vir}$ in these halos as a function of redshift. 
As the Figure shows, the UV metal line luminosities, star formation rates, and mass outflow rates all exhibit strong and correlated time variability. 
In particular, epochs of peak metal line emission coincide closely with massive outflow events in the halos (though some outflow events do not result in strongly enhanced UV line emission), with peak-to-trough variations of up to $\sim2$ dex in luminosity. 
It is noteworthy that peaks in UV line luminosities correspond more closely with peaks in mass outflow rate than with peaks in star formation rate. 
In the FIRE simulations, star formation rate peaks typically precede mass outflow rate (and UV line luminosity) by $\approx 60$ Myr, corresponding to the travel time from the galaxy to $0.25R_{\rm vir}$ \citep[][]{2015MNRAS.454.2691M}. 
This indicates that while metal UV line emission time variability is ultimately driven by star formation time variability, it is the star formation-driven galactic winds specifically that inject energy into the CGM, which is ultimately radiated away by UV metal lines (among other channels). 
This finding is consistent with our conclusion in the previous section that the UV metal line emission arises primarily in collisionally ionized gas, with little sensitivity to the ionizing flux. 
The kinetic energy carried by an outflow with mass outflow rate $\dot{M}_{\rm out}$ and velocity $v_{\rm w}$ is
\begin{align}
\dot{E}_{\rm w} & \equiv \frac{1}{2} \dot{M}_{\rm out} v_{\rm w}^{2} \\
& \approx {\rm 3\times10^{42}~erg~s^{-1}}\left( \frac{\dot{M}_{\rm out}}{\rm 100~M_{\odot}~yr^{-1}} \right) \left( \frac{v_{\rm w}}{\rm 300~km~s^{-1}} \right)^{1/2},
\end{align}
which is sufficient to power the predicted UV metal line emission (see Muratov et al. 2015\nocite{2015MNRAS.454.2691M} for measurements of outflow velocities in our simulations). 
Figure \ref{time_evol} explains why halo {\bf z2h506} is much more luminous than halo {\bf z2h450} at $z=4$ in spite of being nearly identical in mass: at $z\sim4$, {\bf z2h506} experiences a very massive and energetic mass outflow. 

Figure \ref{fig:correlations} shows more quantitatively the correlations between CGM UV metal line luminosities (for the CIII and SiIII lines, which are generally most luminous), star formation rate, and mass outflow rate. 
Overall, CGM UV line luminosities correlate positively with mass outflow rate at fixed redshift but no significant correlation with instantaneous star formation rate is apparent.  
The relationship with instantaneous star formation rate is much weaker because of the time delay between star formation bursts and outflows through $0.25R_{\rm vir}$. 
This again indicates that energy injection from galactic winds is the primary driver of the emission. 
Despite the correlation, the relationship between CGM luminosities and outflow rate is not one-to-one (in some redshift intervals, the relationship even appears inverted) because the luminosity is not simply a function of mass outflow rate but also of how the wind energy is dissipated as the wind interacts with other CGM gas. 
At high redshift, the FIRE simulated halos are very dynamic, which introduces significant fluctuations in how the wind energy is dissipated. 
The FIRE simulations predict that star formation rates and outflows both become more time steady at low redshift ($z\lesssim1$; Muratov et al. 2015), so a tighter relationship between CGM luminosities and star formation rate may result at low redshift. 
Since the metal line emissivity scales with gas metallicity (eq. \ref{eq:Lpart}), another potential source of time variability is fluctuations in gas metallicity. However, the average wind metallicity in our simulations fluctuates by a factor of only $\sim2$ in our simulations (Muratov et al. in prep.) so this must be a subdominant effect on the total order-of-magnitude time variability.

There is some support from observations that star formation-driven outflows inject energy into galaxy halos that can collisionally excite metal lines. 
\cite{2015MNRAS.450.2067T} inferred based on ionization modeling of O VI absorption lines coincident with low HI columns the presence of a substantial mass of metal-enriched, $T>10^{5}$ K collisionally ionized gas around $z\sim2.3$ star-forming galaxies, which they identified with hot outflows from the galaxies. 

An important implication of the star formation-driven time variability of UV metal lines is that CGM gas could be detected in emission in relatively low-mass halos which are experiencing strong outbursts. 
Conversely, detections of CGM emission will in general be biased toward halos that have recently experienced a burst of star formation followed by an energetic outflow. \

\section{Conclusions} \label{s:conclusions}
Motivated by current and upcoming integral field spectrographs on 8-10 m-class telescopes, 
such as MUSE on the VLT and KCWI on Keck,  we used cosmological zoom-in simulations from the FIRE project to make predictions for UV metal line emission from the CGM of $z=2-4$ star-forming galaxies. 
These simulations resolve the ISM of individual galaxies and implement a comprehensive model of stellar feedback, including photoionization, radiation pressure on dust grains, stellar winds, and supernovae of Types I \& II. 
We analyzed 13 simulations with main halos in the mass range $M_{\rm h} \sim 2\times10^{11} - 8.5\times10^{12}$ M$_\odot$ at $z = 2$, representative of Lyman break galaxies. 
For each simulation, we predicted the emission from the \ciii\ (977 \AA), \civ\ (1548 \AA), \siiii\ (1207 \AA), \siiv\ (1394 \AA), \ovi\ (1032 \AA), and \nv\ (1239 \AA) metal lines. 

Our results can be summarized as follows:

\begin{enumerate}
\item Of the transitions we analyze, the low-ionization C III (977 \AA) and Si III (1207 \AA) emission lines are the most luminous, with C IV (1548 \AA) and Si IV (1394 \AA) also showing interesting spatially-extended structures that should be detectable by MUSE and KCWI. 
At $z\leq3$, the CIII and SiIII lines are attenuated by Lyman-series opacity from the intergalactic medium by factors $\lesssim2$, but at $z=2$ the SiIII line is expected to be attenuated by a factor $\sim5$. 
 The more massive halos are on average more UV-luminous. 
\item The UV metal line emission from galactic halos in our simulations arises primarily from collisionally ionized gas and is thus weakly sensitive to the ionizing flux, including ionization from local galaxies.
\item 
The UV metal line emission from galactic halos in our simulations is strongly time variable, with peak-to-trough variations of up to $\sim2$ dex. 
The time variability of UV metal line emission is driven by the time variable star formation predicted by our resolved ISM simulations, but is most closely correlated with the time variability of gas mass outflow rates measured at 0.25$R_{\rm vir}$. 
Our simulations thus indicate that stellar feedback powers UV metal line emission through energy injected in halos by galactic winds.
\end{enumerate}

The prediction that UV metal line emission arises in gas collisionally excited in outflows could be tested by comparing the emission line kinematics with outflow velocities predicted by the simulations. In the FIRE simulations, the median outflow velocity at $0.25R_{\rm vir}$ scales with the halo circular velocity $v_{\rm c} $ \citep[][]{2015MNRAS.454.2691M}. 
On its own, this test may prove ambiguous since $\sim v_{\rm c}$ is also the velocity expected of gas falling into halos. Furthermore, kinematic measurements transverse to galaxies cannot in general distinguish inflows and outflows since it is not known whether the emitting gas is in front or behind the galaxy. 
On the other hand, ``down-the-barrel'' absorption spectroscopy of galaxies \citep[e.g.,][]{2003ApJ...588...65S, ses10, 2012ApJ...760..127M, 2014ApJ...794..156R} can unambiguously probe outflows because the absorbing material is known to lie in front of the stars. 
Thus, correlating outflow kinematics probed by down-the-barrel absorption spectroscopy with circum-galactic UV metal line emission could help identify the UV metal line emission with outflows (with the caveat that a significant time delay may separate peaks in CGM emission and absorption, as well the possible collimation of outflows).

In terms of overall luminosity, our average predictions for UV metal lines from the CGM of high-redshift galaxies are broadly consistent with those of \citet[][]{vs13}, which were based on the large-volume OWLS simulations. 
However, the resolved ISM and stellar feedback physics implemented in the FIRE simulations is critical to capture the strong time variability of star formation and the resulting time variability in UV metal line emission. 
The strong predicted time variability of UV metal line emission has important implications for upcoming observations: even some relatively low-mass halos may be detectable in deep observations with current generation instruments. 
Conversely, flux-limited samples will be biased toward halos whose central galaxy has recently experienced a strong burst of star formation. 

Our finding that galactic winds power UV metal line emission from CGM gas is reminiscent of the ``super wind'' model for Ly$\alpha$ blobs \citep[][]{2000ApJ...532L..13T, 2001ApJ...562L..15T} and highlights the potential of detecting halo gas in emission for constraining stellar feedback. 
Our current calculations do not include the effects of active galactic nuclei (AGN), whose radiative and mechanical interactions with halo gas also likely contribute substantially to metal UV line emission. 
It is clearly worthwhile to include the effects of AGN in future calculations.

\section*{Acknowledgments}
We are grateful to Alex Richings for assistance with the local source test. 
The simulations analyzed in this paper were run on XSEDE computational resources (allocations TG-AST120025, TG-AST130039, and TG-AST140023), on the NASA Pleiades cluster (allocation SMD-14-5189), on the Caltech compute cluster ``Zwicky'' (NSF MRI award \#PHY-0960291), and on the Quest cluster at Northwestern University. 
NS and CAFG were supported by NSF through grants AST-1412836 and AST-1517491, by NASA through grant NNX15AB22G, and by Northwestern University funds. 
DK was supported by NSF grant AST-1412153 and funds from the University of California, San Diego. 
Support for PFH was provided by an Alfred P. Sloan Research Fellowship, NASA ATP grant NNX14AH35G, and NSF Collaborative Research grant \#1411920 and CAREER grant \#1455342. 
RF acknowledges support for this work by NASA through Hubble Fellowship grant HF-51304.01-A awarded by the Space Telescope Science Institute, which is operated by the Association of Universities for Research in Astronomy, Inc., for NASA, under contract NAS 5-26555. 
EQ was supported by NASA ATP grant 12-APT12- 0183, a Simons Investigator award from the Simons Foundation, the David and Lucile Packard Foundation, and the Thomas Alison Schneider Chair in Physics at UC Berkeley.

\bibliography{references}

\begin{thebibliography}{}

\bibitem[\protect\citeauthoryear{{Adelberger}, {Shapley}, {Steidel}, {Pettini},
  {Erb} \& {Reddy}}{{Adelberger} et~al.}{2005}]{2005ApJ...629..636A}
{Adelberger} K.~L.,  {Shapley} A.~E.,  {Steidel} C.~C.,  {Pettini} M.,  {Erb}
  D.~K.,    {Reddy} N.~A.,  2005, \apj, 629, 636

\bibitem[\protect\citeauthoryear{{Adelberger}, {Steidel}, {Pettini}, {Shapley},
  {Reddy} \& {Erb}}{{Adelberger} et~al.}{2005}]{2005ApJ...619..697A}
{Adelberger} K.~L.,  {Steidel} C.~C.,  {Pettini} M.,  {Shapley} A.~E.,  {Reddy}
  N.~A.,    {Erb} D.~K.,  2005, \apj, 619, 697

\bibitem[\protect\citeauthoryear{{Agertz} \& {Kravtsov}}{{Agertz} \&
  {Kravtsov}}{2014}]{2014arXiv1404.2613A}
{Agertz} O.,  {Kravtsov} A.~V.,  2014, ArXiv e-prints

\bibitem[\protect\citeauthoryear{{Agertz}, {Moore}, {Stadel}, {Potter},
  {Miniati}, {Read}, {Mayer}, {Gawryszczak}, {Kravtsov}, {Nordlund}, {Pearce},
  {Quilis}, {Rudd}, {Springel}, {Stone}, {Tasker}, {Teyssier}, {Wadsley} \&
  {Walder}}{{Agertz} et~al.}{2007}]{2007MNRAS.380..963A}
{Agertz} O.,  {Moore} B.,  {Stadel} J.,  {Potter} D.,  {Miniati} F.,  {Read}
  J.,  {Mayer} L.,  {Gawryszczak} A.,  {Kravtsov} A.,  {Nordlund} {\AA}.,
  {Pearce} F.,  {Quilis} V.,  {Rudd} D.,  {Springel} V.,  {Stone} J.,  {Tasker}
  E.,  {Teyssier} R.,  {Wadsley} J.,    {Walder} R.,  2007, \mnras, 380, 963

\bibitem[\protect\citeauthoryear{{Aguirre}, {Hernquist}, {Schaye}, {Weinberg},
  {Katz} \& {Gardner}}{{Aguirre} et~al.}{2001}]{ahs01}
{Aguirre} A.,  {Hernquist} L.,  {Schaye} J.,  {Weinberg} D.~H.,  {Katz} N.,
  {Gardner} J.,  2001, \apj, 560, 599

\bibitem[\protect\citeauthoryear{{Bacon et al.}}{{Bacon et
  al.}}{2010}]{2010SPIE.7735E...7B}
{Bacon et al.} 2010, in Society of Photo-Optical Instrumentation Engineers
  (SPIE) Conference Series Vol.~7735 of Society of Photo-Optical
  Instrumentation Engineers (SPIE) Conference Series, {The MUSE
  second-generation VLT instrument}

\bibitem[\protect\citeauthoryear{{Bacon et al.}}{{Bacon et
  al.}}{2015}]{2015A&A...575A..75B}
{Bacon et al.} 2015, \aap, 575, A75

\bibitem[\protect\citeauthoryear{{Barnes}}{{Barnes}}{2012}]{b12}
{Barnes} J.~E.,  2012, \mnras, 425, 1104

\bibitem[\protect\citeauthoryear{{Bauermeister}, {Blitz} \&
  {Ma}}{{Bauermeister} et~al.}{2010}]{blm10}
{Bauermeister} A.,  {Blitz} L.,    {Ma} C.-P.,  2010, \apj, 717, 323

\bibitem[\protect\citeauthoryear{{Bertone}, {Aguirre} \& {Schaye}}{{Bertone}
  et~al.}{2013}]{bas13}
{Bertone} S.,  {Aguirre} A.,    {Schaye} J.,  2013, \mnras, 430, 3292

\bibitem[\protect\citeauthoryear{{Bertone} \& {Schaye}}{{Bertone} \&
  {Schaye}}{2012}]{bs12}
{Bertone} S.,  {Schaye} J.,  2012, \mnras, 419, 780

\bibitem[\protect\citeauthoryear{{Bertone et al.}}{{Bertone et
  al.}}{2010}]{bsb10}
{Bertone et al.} 2010, \mnras, 408, 1120

\bibitem[\protect\citeauthoryear{{Bouwens}, {Illingworth}, {Franx} \&
  {Ford}}{{Bouwens} et~al.}{2007}]{2007ApJ...670..928B}
{Bouwens} R.~J.,  {Illingworth} G.~D.,  {Franx} M.,    {Ford} H.,  2007, \apj,
  670, 928

\bibitem[\protect\citeauthoryear{{Bryan} \& {Norman}}{{Bryan} \&
  {Norman}}{1998}]{1998ApJ...495...80B}
{Bryan} G.~L.,  {Norman} M.~L.,  1998, \apj, 495, 80

\bibitem[\protect\citeauthoryear{{Cantalupo}}{{Cantalupo}}{2010}]{2010MNRAS.403L..16C}
{Cantalupo} S.,  2010, \mnras, 403, L16

\bibitem[\protect\citeauthoryear{{Cantalupo}, {Arrigoni-Battaia}, {Prochaska},
  {Hennawi} \& {Madau}}{{Cantalupo} et~al.}{2014}]{2014Natur.506...63C}
{Cantalupo} S.,  {Arrigoni-Battaia} F.,  {Prochaska} J.~X.,  {Hennawi} J.~F.,
   {Madau} P.,  2014, \nat, 506, 63

\bibitem[\protect\citeauthoryear{{Cantalupo}, {Porciani}, {Lilly} \&
  {Miniati}}{{Cantalupo} et~al.}{2005}]{cpl05}
{Cantalupo} S.,  {Porciani} C.,  {Lilly} S.~J.,    {Miniati} F.,  2005, \apj,
  628, 61

\bibitem[\protect\citeauthoryear{{Cervi{\~n}o}, {Mas-Hesse} \&
  {Kunth}}{{Cervi{\~n}o} et~al.}{2002}]{2002A&A...392...19C}
{Cervi{\~n}o} M.,  {Mas-Hesse} J.~M.,    {Kunth} D.,  2002, \aap, 392, 19

\bibitem[\protect\citeauthoryear{{Chan}, {Kere{\v s}}, {O{\~n}orbe}, {Hopkins},
  {Muratov}, {Faucher-Gigu{\`e}re} \& {Quataert}}{{Chan}
  et~al.}{2015}]{2015arXiv150702282C}
{Chan} T.~K.,  {Kere{\v s}} D.,  {O{\~n}orbe} J.,  {Hopkins} P.~F.,  {Muratov}
  A.~L.,  {Faucher-Gigu{\`e}re} C.-A.,    {Quataert} E.,  2015, ArXiv e-prints

\bibitem[\protect\citeauthoryear{{Cullen} \& {Dehnen}}{{Cullen} \&
  {Dehnen}}{2010}]{2010MNRAS.408..669C}
{Cullen} L.,  {Dehnen} W.,  2010, \mnras, 408, 669

\bibitem[\protect\citeauthoryear{{Dav{\'e}}, {Katz}, {Oppenheimer}, {Kollmeier}
  \& {Weinberg}}{{Dav{\'e}} et~al.}{2013}]{2013MNRAS.434.2645D}
{Dav{\'e}} R.,  {Katz} N.,  {Oppenheimer} B.~D.,  {Kollmeier} J.~A.,
  {Weinberg} D.~H.,  2013, \mnras, 434, 2645

\bibitem[\protect\citeauthoryear{{Dehnen} \& {Aly}}{{Dehnen} \&
  {Aly}}{2012}]{2012MNRAS.425.1068D}
{Dehnen} W.,  {Aly} H.,  2012, \mnras, 425, 1068

\bibitem[\protect\citeauthoryear{{Dijkstra}, {Haiman} \& {Spaans}}{{Dijkstra}
  et~al.}{2006}]{dhs06}
{Dijkstra} M.,  {Haiman} Z.,    {Spaans} M.,  2006, \apj, 649, 14

\bibitem[\protect\citeauthoryear{{Dijkstra} \& {Loeb}}{{Dijkstra} \&
  {Loeb}}{2009}]{dl09}
{Dijkstra} M.,  {Loeb} A.,  2009, \mnras, 400, 1109

\bibitem[\protect\citeauthoryear{{Durier} \& {Dalla Vecchia}}{{Durier} \&
  {Dalla Vecchia}}{2012}]{2012MNRAS.419..465D}
{Durier} F.,  {Dalla Vecchia} C.,  2012, \mnras, 419, 465

\bibitem[\protect\citeauthoryear{{Fabjan}, {Borgani}, {Tornatore}, {Saro},
  {Murante} \& {Dolag}}{{Fabjan} et~al.}{2010}]{fbt10}
{Fabjan} D.,  {Borgani} S.,  {Tornatore} L.,  {Saro} A.,  {Murante} G.,
  {Dolag} K.,  2010, \mnras, 401, 1670

\bibitem[\protect\citeauthoryear{{Fardal}, {Katz}, {Gardner}, {Hernquist},
  {Weinberg} \& {Dav{\'e}}}{{Fardal} et~al.}{2001}]{fkg01}
{Fardal} M.~A.,  {Katz} N.,  {Gardner} J.~P.,  {Hernquist} L.,  {Weinberg}
  D.~H.,    {Dav{\'e}} R.,  2001, \apj, 562, 605

\bibitem[\protect\citeauthoryear{{Faucher-Gigu\`ere}, {Feldmann}, {Quataert},
  {Keres}, {Hopkins} \& {Murray}}{{Faucher-Gigu\`ere}
  et~al.}{2016}]{2016arXiv160107188F}
{Faucher-Gigu\`ere} C.-A.,  {Feldmann} R.,  {Quataert} E.,  {Keres} D.,
  {Hopkins} P.~F.,    {Murray} N.,  2016, arXiv:1601.07188

\bibitem[\protect\citeauthoryear{{Faucher-Gigu{\`e}re}, {Hopkins}, {Kere{\v
  s}}, {Muratov}, {Quataert} \& {Murray}}{{Faucher-Gigu{\`e}re}
  et~al.}{2015}]{2015MNRAS.449..987F}
{Faucher-Gigu{\`e}re} C.-A.,  {Hopkins} P.~F.,  {Kere{\v s}} D.,  {Muratov}
  A.~L.,  {Quataert} E.,    {Murray} N.,  2015, \mnras, 449, 987

\bibitem[\protect\citeauthoryear{{Faucher-Gigu{\`e}re} \& {Kere{\v
  s}}}{{Faucher-Gigu{\`e}re} \& {Kere{\v s}}}{2011}]{2011MNRAS.412L.118F}
{Faucher-Gigu{\`e}re} C.-A.,  {Kere{\v s}} D.,  2011, \mnras, 412, L118

\bibitem[\protect\citeauthoryear{{Faucher-Gigu{\`e}re}, {Kere{\v s}},
  {Dijkstra}, {Hernquist} \& {Zaldarriaga}}{{Faucher-Gigu{\`e}re}
  et~al.}{2010}]{fkd10}
{Faucher-Gigu{\`e}re} C.-A.,  {Kere{\v s}} D.,  {Dijkstra} M.,  {Hernquist} L.,
     {Zaldarriaga} M.,  2010, \apj, 725, 633

\bibitem[\protect\citeauthoryear{{Faucher-Gigu{\`e}re}, {Kere{\v s}} \&
  {Ma}}{{Faucher-Gigu{\`e}re} et~al.}{2011}]{fkm11}
{Faucher-Gigu{\`e}re} C.-A.,  {Kere{\v s}} D.,    {Ma} C.-P.,  2011, \mnras,
  417, 2982

\bibitem[\protect\citeauthoryear{{Faucher-Gigu{\`e}re}, {Lidz}, {Zaldarriaga}
  \& {Hernquist}}{{Faucher-Gigu{\`e}re} et~al.}{2009}]{2009ApJ...703.1416F}
{Faucher-Gigu{\`e}re} C.-A.,  {Lidz} A.,  {Zaldarriaga} M.,    {Hernquist} L.,
  2009, \apj, 703, 1416

\bibitem[\protect\citeauthoryear{{Ferland}, {Porter}, {van Hoof}, {Williams},
  {Abel}, {Lykins}, {Shaw}, {Henney} \& {Stancil}}{{Ferland}
  et~al.}{2013}]{fpv13}
{Ferland} G.~J.,  {Porter} R.~L.,  {van Hoof} P.~A.~M.,  {Williams} R.~J.~R.,
  {Abel} N.~P.,  {Lykins} M.~L.,  {Shaw} G.,  {Henney} W.~J.,    {Stancil}
  P.~C.,  2013, \rmxaa, 49, 137

\bibitem[\protect\citeauthoryear{{F{\"o}rster Schreiber et al.}}{{F{\"o}rster
  Schreiber et al.}}{2009}]{2009ApJ...706.1364F}
{F{\"o}rster Schreiber et al.} 2009, \apj, 706, 1364

\bibitem[\protect\citeauthoryear{{Frank}, {Rasera}, {Vibert}, {Milliard},
  {Popping}, {Blaizot}, {Courty}, {Deharveng}, {P{\'e}roux}, {Teyssier} \&
  {Martin}}{{Frank} et~al.}{2012}]{frv12}
{Frank} S.,  {Rasera} Y.,  {Vibert} D.,  {Milliard} B.,  {Popping} A.,
  {Blaizot} J.,  {Courty} S.,  {Deharveng} J.-M.,  {P{\'e}roux} C.,  {Teyssier}
  R.,    {Martin} C.~D.,  2012, \mnras, 420, 1731

\bibitem[\protect\citeauthoryear{{Fumagalli}, {Fossati}, {Hau}, {Gavazzi},
  {Bower}, {Sun} \& {Boselli}}{{Fumagalli} et~al.}{2014}]{2014MNRAS.445.4335F}
{Fumagalli} M.,  {Fossati} M.,  {Hau} G.~K.~T.,  {Gavazzi} G.,  {Bower} R.,
  {Sun} M.,    {Boselli} A.,  2014, \mnras, 445, 4335

\bibitem[\protect\citeauthoryear{{Furlanetto}, {Schaye}, {Springel} \&
  {Hernquist}}{{Furlanetto} et~al.}{2004}]{fss04}
{Furlanetto} S.~R.,  {Schaye} J.,  {Springel} V.,    {Hernquist} L.,  2004,
  \apj, 606, 221

\bibitem[\protect\citeauthoryear{{Geach et al.}}{{Geach et
  al.}}{2009}]{2009ApJ...700....1G}
{Geach et al.} 2009, \apj, 700, 1

\bibitem[\protect\citeauthoryear{{Geach et al.}}{{Geach et
  al.}}{2014}]{2014ApJ...793...22G}
{Geach et al.} 2014, \apj, 793, 22

\bibitem[\protect\citeauthoryear{{Genzel et al.}}{{Genzel et
  al.}}{2008}]{2008ApJ...687...59G}
{Genzel et al.} 2008, \apj, 687, 59

\bibitem[\protect\citeauthoryear{{Gnat} \& {Sternberg}}{{Gnat} \&
  {Sternberg}}{2007}]{2007ApJS..168..213G}
{Gnat} O.,  {Sternberg} A.,  2007, \apjs, 168, 213

\bibitem[\protect\citeauthoryear{{Goerdt}, {Dekel}, {Sternberg}, {Ceverino},
  {Teyssier} \& {Primack}}{{Goerdt} et~al.}{2010}]{2010MNRAS.407..613G}
{Goerdt} T.,  {Dekel} A.,  {Sternberg} A.,  {Ceverino} D.,  {Teyssier} R.,
  {Primack} J.~R.,  2010, \mnras, 407, 613

\bibitem[\protect\citeauthoryear{{Gould} \& {Weinberg}}{{Gould} \&
  {Weinberg}}{1996}]{1996ApJ...468..462G}
{Gould} A.,  {Weinberg} D.~H.,  1996, \apj, 468, 462

\bibitem[\protect\citeauthoryear{{Governato}, {Brook}, {Mayer}, {Brooks},
  {Rhee}, {Wadsley}, {Jonsson}, {Willman}, {Stinson}, {Quinn} \&
  {Madau}}{{Governato} et~al.}{2010}]{2010Natur.463..203G}
{Governato} F.,  {Brook} C.,  {Mayer} L.,  {Brooks} A.,  {Rhee} G.,  {Wadsley}
  J.,  {Jonsson} P.,  {Willman} B.,  {Stinson} G.,  {Quinn} T.,    {Madau} P.,
  2010, \nat, 463, 203

\bibitem[\protect\citeauthoryear{{Guedes}, {Callegari}, {Madau} \&
  {Mayer}}{{Guedes} et~al.}{2011}]{2011ApJ...742...76G}
{Guedes} J.,  {Callegari} S.,  {Madau} P.,    {Mayer} L.,  2011, \apj, 742, 76

\bibitem[\protect\citeauthoryear{{Haiman}, {Spaans} \& {Quataert}}{{Haiman}
  et~al.}{2000}]{2000ApJ...537L...5H}
{Haiman} Z.,  {Spaans} M.,    {Quataert} E.,  2000, \apjl, 537, L5

\bibitem[\protect\citeauthoryear{{Hennawi}, {Prochaska}, {Kollmeier} \&
  {Zheng}}{{Hennawi} et~al.}{2009}]{2009ApJ...693L..49H}
{Hennawi} J.~F.,  {Prochaska} J.~X.,  {Kollmeier} J.,    {Zheng} Z.,  2009,
  \apjl, 693, L49

\bibitem[\protect\citeauthoryear{{Hopkins}}{{Hopkins}}{2013}]{2013MNRAS.428.2840H}
{Hopkins} P.~F.,  2013, \mnras, 428, 2840

\bibitem[\protect\citeauthoryear{{Hopkins}}{{Hopkins}}{2014}]{2014arXiv1409.7395H}
{Hopkins} P.~F.,  2014, ArXiv e-prints

\bibitem[\protect\citeauthoryear{{Hopkins}, {Kere{\v s}}, {O{\~n}orbe},
  {Faucher-Gigu{\`e}re}, {Quataert}, {Murray} \& {Bullock}}{{Hopkins}
  et~al.}{2014}]{2014MNRAS.445..581H}
{Hopkins} P.~F.,  {Kere{\v s}} D.,  {O{\~n}orbe} J.,  {Faucher-Gigu{\`e}re}
  C.-A.,  {Quataert} E.,  {Murray} N.,    {Bullock} J.~S.,  2014, \mnras, 445,
  581

\bibitem[\protect\citeauthoryear{{Inoue}, {Shimizu}, {Iwata} \&
  {Tanaka}}{{Inoue} et~al.}{2014}]{2014MNRAS.442.1805I}
{Inoue} A.~K.,  {Shimizu} I.,  {Iwata} I.,    {Tanaka} M.,  2014, \mnras, 442,
  1805

\bibitem[\protect\citeauthoryear{{Kere{\v s}} \& {Hernquist}}{{Kere{\v s}} \&
  {Hernquist}}{2009}]{2009ApJ...700L...1K}
{Kere{\v s}} D.,  {Hernquist} L.,  2009, \apjl, 700, L1

\bibitem[\protect\citeauthoryear{{Kere{\v s}}, {Katz}, {Weinberg} \&
  {Dav{\'e}}}{{Kere{\v s}} et~al.}{2005}]{kk05}
{Kere{\v s}} D.,  {Katz} N.,  {Weinberg} D.~H.,    {Dav{\'e}} R.,  2005,
  \mnras, 363, 2

\bibitem[\protect\citeauthoryear{{Kim et al.}}{{Kim et
  al.}}{2014}]{2014ApJS..210...14K}
{Kim et al.} 2014, \apjs, 210, 14

\bibitem[\protect\citeauthoryear{{Knollmann} \& {Knebe}}{{Knollmann} \&
  {Knebe}}{2009}]{2009ApJS..182..608K}
{Knollmann} S.~R.,  {Knebe} A.,  2009, \apjs, 182, 608

\bibitem[\protect\citeauthoryear{{Kollmeier}, {Zheng}, {Dav{\'e}}, {Gould},
  {Katz}, {Miralda-Escud{\'e}} \& {Weinberg}}{{Kollmeier}
  et~al.}{2010}]{2010ApJ...708.1048K}
{Kollmeier} J.~A.,  {Zheng} Z.,  {Dav{\'e}} R.,  {Gould} A.,  {Katz} N.,
  {Miralda-Escud{\'e}} J.,    {Weinberg} D.~H.,  2010, \apj, 708, 1048

\bibitem[\protect\citeauthoryear{{Law}, {Steidel}, {Erb}, {Larkin}, {Pettini},
  {Shapley} \& {Wright}}{{Law} et~al.}{2009}]{2009ApJ...697.2057L}
{Law} D.~R.,  {Steidel} C.~C.,  {Erb} D.~K.,  {Larkin} J.~E.,  {Pettini} M.,
  {Shapley} A.~E.,    {Wright} S.~A.,  2009, \apj, 697, 2057

\bibitem[\protect\citeauthoryear{{Leitherer}, {Schaerer}, {Goldader},
  {Delgado}, {Robert}, {Kune}, {de Mello}, {Devost} \& {Heckman}}{{Leitherer}
  et~al.}{1999}]{1999ApJS..123....3L}
{Leitherer} C.,  {Schaerer} D.,  {Goldader} J.~D.,  {Delgado} R.~M.~G.,
  {Robert} C.,  {Kune} D.~F.,  {de Mello} D.~F.,  {Devost} D.,    {Heckman}
  T.~M.,  1999, \apjs, 123, 3

\bibitem[\protect\citeauthoryear{{Ma}, {Hopkins}, {Faucher-Gigu{\`e}re},
  {Zolman}, {Muratov}, {Kere{\v s}} \& {Quataert}}{{Ma}
  et~al.}{2016}]{2016MNRAS.456.2140M}
{Ma} X.,  {Hopkins} P.~F.,  {Faucher-Gigu{\`e}re} C.-A.,  {Zolman} N.,
  {Muratov} A.~L.,  {Kere{\v s}} D.,    {Quataert} E.,  2016, \mnras, 456, 2140

\bibitem[\protect\citeauthoryear{{Madau}}{{Madau}}{1995}]{1995ApJ...441...18M}
{Madau} P.,  1995, \apj, 441, 18

\bibitem[\protect\citeauthoryear{{Martin}, {Moore}, {Morrissey}, {Matuszewski},
  {Rahman}, {Adkins} \& {Epps}}{{Martin} et~al.}{2010}]{2010SPIE.7735E..21M}
{Martin} C.,  {Moore} A.,  {Morrissey} P.,  {Matuszewski} M.,  {Rahman} S.,
  {Adkins} S.,    {Epps} H.,  2010, in Society of Photo-Optical Instrumentation
  Engineers (SPIE) Conference Series Vol.~7735 of Society of Photo-Optical
  Instrumentation Engineers (SPIE) Conference Series, {The Keck Cosmic Web
  Imager}

\bibitem[\protect\citeauthoryear{{Martin}, {Scannapieco}, {Ellison}, {Hennawi},
  {Djorgovski} \& {Fournier}}{{Martin} et~al.}{2010}]{mse10}
{Martin} C.~L.,  {Scannapieco} E.,  {Ellison} S.~L.,  {Hennawi} J.~F.,
  {Djorgovski} S.~G.,    {Fournier} A.~P.,  2010, \apj, 721, 174

\bibitem[\protect\citeauthoryear{{Martin}, {Shapley}, {Coil}, {Kornei},
  {Bundy}, {Weiner}, {Noeske} \& {Schiminovich}}{{Martin}
  et~al.}{2012}]{2012ApJ...760..127M}
{Martin} C.~L.,  {Shapley} A.~E.,  {Coil} A.~L.,  {Kornei} K.~A.,  {Bundy} K.,
  {Weiner} B.~J.,  {Noeske} K.~G.,    {Schiminovich} D.,  2012, \apj, 760, 127

\bibitem[\protect\citeauthoryear{{Martin}, {Chang}, {Matuszewski}, {Morrissey},
  {Rahman}, {Moore} \& {Steidel}}{{Martin} et~al.}{2014}]{2014ApJ...786..106M}
{Martin} D.~C.,  {Chang} D.,  {Matuszewski} M.,  {Morrissey} P.,  {Rahman} S.,
  {Moore} A.,    {Steidel} C.~C.,  2014, \apj, 786, 106

\bibitem[\protect\citeauthoryear{{Martin et al.}}{{Martin et
  al.}}{2014a}]{mcm14a}
{Martin et al.} 2014a, \apj, 786, 106

\bibitem[\protect\citeauthoryear{{Martin et al.}}{{Martin et
  al.}}{2014b}]{mcm14b}
{Martin et al.} 2014b, \apj, 786, 107

\bibitem[\protect\citeauthoryear{{Mas-Hesse}, {Ot{\'{\i}}-Floranes} \&
  {Cervi{\~n}o}}{{Mas-Hesse} et~al.}{2008}]{2008A&A...483...71M}
{Mas-Hesse} J.~M.,  {Ot{\'{\i}}-Floranes} H.,    {Cervi{\~n}o} M.,  2008, \aap,
  483, 71

\bibitem[\protect\citeauthoryear{{Matsuda}, {Yamada}, {Hayashino}, {Tamura},
  {Yamauchi}, {Ajiki}, {Fujita}, {Murayama}, {Nagao}, {Ohta}, {Okamura},
  {Ouchi}, {Shimasaku}, {Shioya} \& {Taniguchi}}{{Matsuda}
  et~al.}{2004}]{2004AJ....128..569M}
{Matsuda} Y.,  {Yamada} T.,  {Hayashino} T.,  {Tamura} H.,  {Yamauchi} R.,
  {Ajiki} M.,  {Fujita} S.~S.,  {Murayama} T.,  {Nagao} T.,  {Ohta} K.,
  {Okamura} S.,  {Ouchi} M.,  {Shimasaku} K.,  {Shioya} Y.,    {Taniguchi} Y.,
  2004, \aj, 128, 569

\bibitem[\protect\citeauthoryear{{Matuszewski}, {Chang}, {Crabill}, {Martin},
  {Moore}, {Morrissey} \& {Rahman}}{{Matuszewski}
  et~al.}{2010}]{2010SPIE.7735E..24M}
{Matuszewski} M.,  {Chang} D.,  {Crabill} R.~M.,  {Martin} D.~C.,  {Moore}
  A.~M.,  {Morrissey} P.,    {Rahman} S.,  2010, in Society of Photo-Optical
  Instrumentation Engineers (SPIE) Conference Series Vol.~7735 of Society of
  Photo-Optical Instrumentation Engineers (SPIE) Conference Series, {The Cosmic
  Web Imager: an integral field spectrograph for the Hale Telescope at Palomar
  Observatory: instrument design and first results}

\bibitem[\protect\citeauthoryear{{Muratov}, {Kere{\v s}},
  {Faucher-Gigu{\`e}re}, {Hopkins}, {Quataert} \& {Murray}}{{Muratov}
  et~al.}{2015}]{2015MNRAS.454.2691M}
{Muratov} A.~L.,  {Kere{\v s}} D.,  {Faucher-Gigu{\`e}re} C.-A.,  {Hopkins}
  P.~F.,  {Quataert} E.,    {Murray} N.,  2015, \mnras, 454, 2691

\bibitem[\protect\citeauthoryear{{O{\~n}orbe}, {Boylan-Kolchin}, {Bullock},
  {Hopkins}, {Ker{\v e}s}, {Faucher-Gigu{\`e}re}, {Quataert} \&
  {Murray}}{{O{\~n}orbe} et~al.}{2015}]{2015arXiv150202036O}
{O{\~n}orbe} J.,  {Boylan-Kolchin} M.,  {Bullock} J.~S.,  {Hopkins} P.~F.,
  {Ker{\v e}s} D.,  {Faucher-Gigu{\`e}re} C.-A.,  {Quataert} E.,    {Murray}
  N.,  2015, ArXiv e-prints

\bibitem[\protect\citeauthoryear{{Oppenheimer} \& {Dav{\'e}}}{{Oppenheimer} \&
  {Dav{\'e}}}{2006}]{od06}
{Oppenheimer} B.~D.,  {Dav{\'e}} R.,  2006, \mnras, 373, 1265

\bibitem[\protect\citeauthoryear{{Oppenheimer} \& {Schaye}}{{Oppenheimer} \&
  {Schaye}}{2013}]{2013MNRAS.434.1043O}
{Oppenheimer} B.~D.,  {Schaye} J.,  2013, \mnras, 434, 1043

\bibitem[\protect\citeauthoryear{{Prescott}, {Momcheva}, {Brammer}, {Fynbo} \&
  {M{\o}ller}}{{Prescott} et~al.}{2015}]{2015arXiv150105312P}
{Prescott} M.~K.~M.,  {Momcheva} I.,  {Brammer} G.~B.,  {Fynbo} J.~P.~U.,
  {M{\o}ller} P.,  2015, ArXiv e-prints

\bibitem[\protect\citeauthoryear{{Price}}{{Price}}{2008}]{2008JCoPh.22710040P}
{Price} D.~J.,  2008, Journal of Computational Physics, 227, 10040

\bibitem[\protect\citeauthoryear{{Price} \& {Monaghan}}{{Price} \&
  {Monaghan}}{2007}]{pm07}
{Price} D.~J.,  {Monaghan} J.~J.,  2007, \mnras, 374, 1347

\bibitem[\protect\citeauthoryear{{Prochaska} \& {Hennawi}}{{Prochaska} \&
  {Hennawi}}{2009}]{2009ApJ...690.1558P}
{Prochaska} J.~X.,  {Hennawi} J.~F.,  2009, \apj, 690, 1558

\bibitem[\protect\citeauthoryear{{Prochaska} \& {Wolfe}}{{Prochaska} \&
  {Wolfe}}{2009}]{pw09}
{Prochaska} J.~X.,  {Wolfe} A.~M.,  2009, \apj, 696, 1543

\bibitem[\protect\citeauthoryear{{Rauch}, {Haehnelt}, {Bunker}, {Becker},
  {Marleau}, {Graham}, {Cristiani}, {Jarvis}, {Lacey}, {Morris}, {Peroux},
  {R{\"o}ttgering} \& {Theuns}}{{Rauch} et~al.}{2008}]{2008ApJ...681..856R}
{Rauch} M.,  {Haehnelt} M.,  {Bunker} A.,  {Becker} G.,  {Marleau} F.,
  {Graham} J.,  {Cristiani} S.,  {Jarvis} M.,  {Lacey} C.,  {Morris} S.,
  {Peroux} C.,  {R{\"o}ttgering} H.,    {Theuns} T.,  2008, \apj, 681, 856

\bibitem[\protect\citeauthoryear{{Richard}, {Patricio}, {Martinez}, {Bacon},
  {Cl{\'e}ment}, {Weilbacher}, {Soto}, {Wisotzki}, {Vernet}, {Pello}, {Schaye},
  {Turner} \& {Martinsson}}{{Richard} et~al.}{2015}]{2015MNRAS.446L..16R}
{Richard} J.,  {Patricio} V.,  {Martinez} J.,  {Bacon} R.,  {Cl{\'e}ment} B.,
  {Weilbacher} P.,  {Soto} K.,  {Wisotzki} L.,  {Vernet} J.,  {Pello} R.,
  {Schaye} J.,  {Turner} M.,    {Martinsson} T.,  2015, \mnras, 446, L16

\bibitem[\protect\citeauthoryear{{Rosdahl} \& {Blaizot}}{{Rosdahl} \&
  {Blaizot}}{2012}]{2012MNRAS.423..344R}
{Rosdahl} J.,  {Blaizot} J.,  2012, \mnras, 423, 344

\bibitem[\protect\citeauthoryear{{Rubin}, {Prochaska}, {Koo}, {Phillips},
  {Martin} \& {Winstrom}}{{Rubin} et~al.}{2014}]{2014ApJ...794..156R}
{Rubin} K.~H.~R.,  {Prochaska} J.~X.,  {Koo} D.~C.,  {Phillips} A.~C.,
  {Martin} C.~L.,    {Winstrom} L.~O.,  2014, \apj, 794, 156

\bibitem[\protect\citeauthoryear{{Rubin}, {Weiner}, {Koo}, {Martin},
  {Prochaska}, {Coil} \& {Newman}}{{Rubin} et~al.}{2010}]{2010ApJ...719.1503R}
{Rubin} K.~H.~R.,  {Weiner} B.~J.,  {Koo} D.~C.,  {Martin} C.~L.,  {Prochaska}
  J.~X.,  {Coil} A.~L.,    {Newman} J.~A.,  2010, \apj, 719, 1503

\bibitem[\protect\citeauthoryear{{Rudie}, {Steidel}, {Trainor}, {Rakic},
  {Bogosavljevi{\'c}}, {Pettini}, {Reddy}, {Shapley}, {Erb} \& {Law}}{{Rudie}
  et~al.}{2012}]{2012ApJ...750...67R}
{Rudie} G.~C.,  {Steidel} C.~C.,  {Trainor} R.~F.,  {Rakic} O.,
  {Bogosavljevi{\'c}} M.,  {Pettini} M.,  {Reddy} N.,  {Shapley} A.~E.,  {Erb}
  D.~K.,    {Law} D.~R.,  2012, \apj, 750, 67

\bibitem[\protect\citeauthoryear{{Schaye et al.}}{{Schaye et
  al.}}{2015}]{2015MNRAS.446..521S}
{Schaye et al.} 2015, \mnras, 446, 521

\bibitem[\protect\citeauthoryear{{Shapley}, {Steidel}, {Pettini} \&
  {Adelberger}}{{Shapley} et~al.}{2003}]{2003ApJ...588...65S}
{Shapley} A.~E.,  {Steidel} C.~C.,  {Pettini} M.,    {Adelberger} K.~L.,  2003,
  \apj, 588, 65

\bibitem[\protect\citeauthoryear{{Shapley}, {Steidel}, {Pettini}, {Adelberger}
  \& {Erb}}{{Shapley} et~al.}{2006}]{2006ApJ...651..688S}
{Shapley} A.~E.,  {Steidel} C.~C.,  {Pettini} M.,  {Adelberger} K.~L.,    {Erb}
  D.~K.,  2006, \apj, 651, 688

\bibitem[\protect\citeauthoryear{{Shen}, {Wadsley} \& {Stinson}}{{Shen}
  et~al.}{2010}]{2010MNRAS.407.1581S}
{Shen} S.,  {Wadsley} J.,    {Stinson} G.,  2010, \mnras, 407, 1581

\bibitem[\protect\citeauthoryear{{Sijacki}, {Vogelsberger}, {Kere{\v s}},
  {Springel} \& {Hernquist}}{{Sijacki} et~al.}{2012}]{2012MNRAS.424.2999S}
{Sijacki} D.,  {Vogelsberger} M.,  {Kere{\v s}} D.,  {Springel} V.,
  {Hernquist} L.,  2012, \mnras, 424, 2999

\bibitem[\protect\citeauthoryear{{Somerville} \& {Dav{\'e}}}{{Somerville} \&
  {Dav{\'e}}}{2014}]{2014arXiv1412.2712S}
{Somerville} R.~S.,  {Dav{\'e}} R.,  2014, ArXiv e-prints

\bibitem[\protect\citeauthoryear{{Sparre}, {Hayward}, {Feldmann},
  {Faucher-Gigu{\`e}re}, {Muratov}, {Kere{\v s}} \& {Hopkins}}{{Sparre}
  et~al.}{2015}]{2015arXiv151003869S}
{Sparre} M.,  {Hayward} C.~C.,  {Feldmann} R.,  {Faucher-Gigu{\`e}re} C.-A.,
  {Muratov} A.~L.,  {Kere{\v s}} D.,    {Hopkins} P.~F.,  2015, ArXiv e-prints

\bibitem[\protect\citeauthoryear{{Springel}}{{Springel}}{2005}]{2005MNRAS.364.1105S}
{Springel} V.,  2005, \mnras, 364, 1105

\bibitem[\protect\citeauthoryear{{Springel} \& {Hernquist}}{{Springel} \&
  {Hernquist}}{2003}]{sh03}
{Springel} V.,  {Hernquist} L.,  2003, \mnras, 339, 289

\bibitem[\protect\citeauthoryear{{Steidel}, {Adelberger}, {Shapley}, {Pettini},
  {Dickinson} \& {Giavalisco}}{{Steidel} et~al.}{2000}]{2000ApJ...532..170S}
{Steidel} C.~C.,  {Adelberger} K.~L.,  {Shapley} A.~E.,  {Pettini} M.,
  {Dickinson} M.,    {Giavalisco} M.,  2000, \apj, 532, 170

\bibitem[\protect\citeauthoryear{{Steidel}, {Bogosavljevi{\'c}}, {Shapley},
  {Kollmeier}, {Reddy}, {Erb} \& {Pettini}}{{Steidel}
  et~al.}{2011}]{2011ApJ...736..160S}
{Steidel} C.~C.,  {Bogosavljevi{\'c}} M.,  {Shapley} A.~E.,  {Kollmeier} J.~A.,
   {Reddy} N.~A.,  {Erb} D.~K.,    {Pettini} M.,  2011, \apj, 736, 160

\bibitem[\protect\citeauthoryear{{Steidel}, {Erb}, {Shapley}, {Pettini},
  {Reddy}, {Bogosavljevi{\'c}}, {Rudie} \& {Rakic}}{{Steidel}
  et~al.}{2010}]{ses10}
{Steidel} C.~C.,  {Erb} D.~K.,  {Shapley} A.~E.,  {Pettini} M.,  {Reddy} N.,
  {Bogosavljevi{\'c}} M.,  {Rudie} G.~C.,    {Rakic} O.,  2010, \apj, 717, 289

\bibitem[\protect\citeauthoryear{{Stinson}, {Brook}, {Macci{\`o}}, {Wadsley},
  {Quinn} \& {Couchman}}{{Stinson} et~al.}{2013}]{2013MNRAS.428..129S}
{Stinson} G.~S.,  {Brook} C.,  {Macci{\`o}} A.~V.,  {Wadsley} J.,  {Quinn}
  T.~R.,    {Couchman} H.~M.~P.,  2013, \mnras, 428, 129

\bibitem[\protect\citeauthoryear{{Taniguchi} \& {Shioya}}{{Taniguchi} \&
  {Shioya}}{2000}]{2000ApJ...532L..13T}
{Taniguchi} Y.,  {Shioya} Y.,  2000, \apjl, 532, L13

\bibitem[\protect\citeauthoryear{{Taniguchi}, {Shioya} \& {Kakazu}}{{Taniguchi}
  et~al.}{2001}]{2001ApJ...562L..15T}
{Taniguchi} Y.,  {Shioya} Y.,    {Kakazu} Y.,  2001, \apjl, 562, L15

\bibitem[\protect\citeauthoryear{{Trainor} \& {Steidel}}{{Trainor} \&
  {Steidel}}{2012}]{2012ApJ...752...39T}
{Trainor} R.~F.,  {Steidel} C.~C.,  2012, \apj, 752, 39

\bibitem[\protect\citeauthoryear{{Tumlinson}, {Thom}, {Werk}, {Prochaska},
  {Tripp}, {Katz}, {Dav{\'e}}, {Oppenheimer}, {Meiring}, {Ford}, {O'Meara},
  {Peeples}, {Sembach} \& {Weinberg}}{{Tumlinson}
  et~al.}{2013}]{2013ApJ...777...59T}
{Tumlinson} J.,  {Thom} C.,  {Werk} J.~K.,  {Prochaska} J.~X.,  {Tripp} T.~M.,
  {Katz} N.,  {Dav{\'e}} R.,  {Oppenheimer} B.~D.,  {Meiring} J.~D.,  {Ford}
  A.~B.,  {O'Meara} J.~M.,  {Peeples} M.~S.,  {Sembach} K.~R.,    {Weinberg}
  D.~H.,  2013, \apj, 777, 59

\bibitem[\protect\citeauthoryear{{Turner}, {Schaye}, {Steidel}, {Rudie} \&
  {Strom}}{{Turner} et~al.}{2015}]{2015MNRAS.450.2067T}
{Turner} M.~L.,  {Schaye} J.,  {Steidel} C.~C.,  {Rudie} G.~C.,    {Strom}
  A.~L.,  2015, \mnras, 450, 2067

\bibitem[\protect\citeauthoryear{{van de Voort} \& {Schaye}}{{van de Voort} \&
  {Schaye}}{2013}]{vs13}
{van de Voort} F.,  {Schaye} J.,  2013, \mnras, 430, 2688

\bibitem[\protect\citeauthoryear{{Verhamme}, {Schaerer} \&
  {Maselli}}{{Verhamme} et~al.}{2006}]{vsm06}
{Verhamme} A.,  {Schaerer} D.,    {Maselli} A.,  2006, \aap, 460, 397

\bibitem[\protect\citeauthoryear{{Vogelsberger}, {Genel}, {Springel}, {Torrey},
  {Sijacki}, {Xu}, {Snyder}, {Nelson} \& {Hernquist}}{{Vogelsberger}
  et~al.}{2014}]{2014MNRAS.444.1518V}
{Vogelsberger} M.,  {Genel} S.,  {Springel} V.,  {Torrey} P.,  {Sijacki} D.,
  {Xu} D.,  {Snyder} G.,  {Nelson} D.,    {Hernquist} L.,  2014, \mnras, 444,
  1518

\bibitem[\protect\citeauthoryear{{Wiersma}, {Schaye} \& {Smith}}{{Wiersma}
  et~al.}{2009}]{2009MNRAS.393...99W}
{Wiersma} R.~P.~C.,  {Schaye} J.,    {Smith} B.~D.,  2009, \mnras, 393, 99

\bibitem[\protect\citeauthoryear{{Wiersma}, {Schaye} \& {Theuns}}{{Wiersma}
  et~al.}{2011}]{wst11}
{Wiersma} R.~P.~C.,  {Schaye} J.,    {Theuns} T.,  2011, \mnras, 415, 353

\bibitem[\protect\citeauthoryear{{Wisotzki et al.}}{{Wisotzki et
  al.}}{2015}]{2015arXiv150905143W}
{Wisotzki et al.} 2015, ArXiv e-prints

\bibitem[\protect\citeauthoryear{{Wright}, {Larkin}, {Law}, {Steidel},
  {Shapley} \& {Erb}}{{Wright} et~al.}{2009}]{2009ApJ...699..421W}
{Wright} S.~A.,  {Larkin} J.~E.,  {Law} D.~R.,  {Steidel} C.~C.,  {Shapley}
  A.~E.,    {Erb} D.~K.,  2009, \apj, 699, 421

\bibitem[\protect\citeauthoryear{{Yang}, {Zabludoff}, {Tremonti}, {Eisenstein}
  \& {Dav{\'e}}}{{Yang} et~al.}{2009}]{2009ApJ...693.1579Y}
{Yang} Y.,  {Zabludoff} A.,  {Tremonti} C.,  {Eisenstein} D.,    {Dav{\'e}} R.,
   2009, \apj, 693, 1579

\end{thebibliography}

\appendix
\section{Convergence Test}
\label{sec:convergence}

\begin{figure*}
\begin{center}
\includegraphics[width=\textwidth, trim = 0.35in 0.0in 0.5in 0.3in, clip=true]{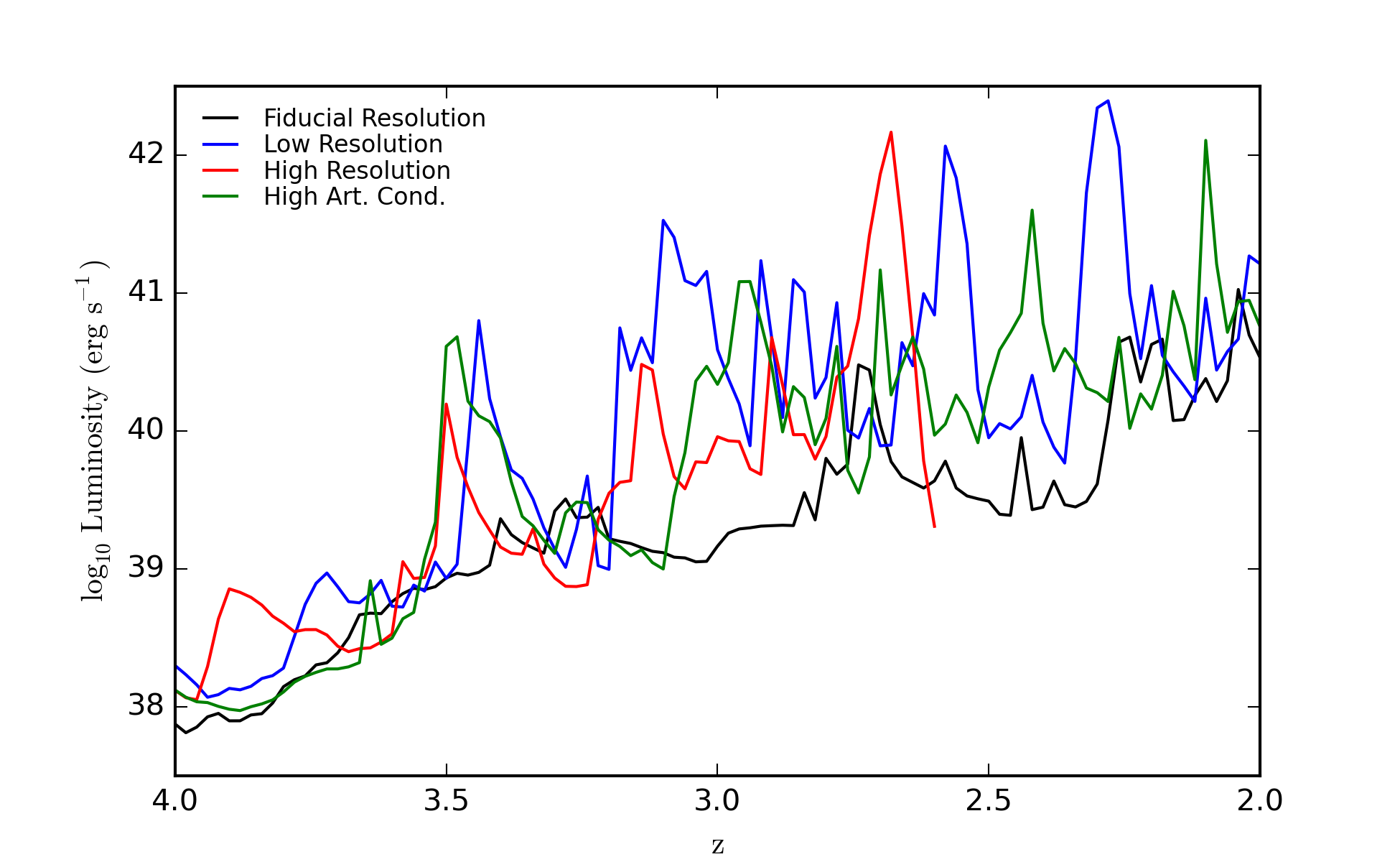}
\end{center}
\caption[]{Convergence test for the predicted \ciii~luminosity for {\bf m12i} with respect to the simulation resolution and the artificial conductivity normalization. 
We plot the \ciii\ luminosity within a virial radius as a function of redshift for the fiducial resolution (analyzed in the main text), for a lower resolution run, for a higher resolution run (stopped at an earlier redshift), and for fiducial resolution run with the normalization of artificial conductivity multiplied by a factor of 4 (see text for more details). 
Accounting for the stochasticity of the simulations, we conclude that the predicted \ciii~luminosity does not depend systematically on resolution or on the normalization of the artificial conductivity, and thus that our predictions are reasonably converged for our fiducial resolution simulations. 
 }
\label{f:convergence} 
\end{figure*}

We test the convergence of our predictions for the intensity of metal emission lines 
with respect to the resolution of the SPH calculation. 
To do this we use three versions of the {\bf m12i} simulation: one having a `fiducial resolution' (the one analyzed in the main text and listed in Table \ref{t:simulations}), one having a `low resolution' (with a gas particle mass 8 times larger and a minimum gas softening length 5 times larger), and one having a `high resolution' (with a gas particle mass 8 times smaller but same minimum softening).  The `high resolution' version was not evolved all the way to $z=2$ due to its high computational cost; the figure shows this simulation only to $z=2.6$. We also use another version of this halo whose parameters are similar to the `fiducial resolution' version but in which the normalization of the artificial conductivity (entropy mixing parameter) is multiplied by a factor of 4. 

In Figure \ref{f:convergence} we plot the \ciii\ luminosity within a virial radius as a function of redshift in all four realizations of the {\bf m12i} halo described above. In this calculation we have excluded a region of 10 proper kpc centered on the halos as a proxy for removing emission from the galaxy. Consistent with the results found in \S~\ref{s:time_var}, the luminosities for all four realizations are highly time-variable. Accounting for the stochasticity of the simulations, we conclude that the predicted \ciii~luminosity does not depend systematically on resolution or on the normalization of the artificial conductivity, and thus that our predictions are reasonably converged for our fiducial resolution simulations. 
 
\section{Local Source Test}
\label{sec:local_xrays}
To further test the sensitivity of our CGM emission predictions to local sources of ionizing radiation, we repeated our calculations explicitly including a local source spectrum in addition to the FG09 cosmic background. 
Specifically, we added a local source spectrum from \cite{2002A&A...392...19C} that includes X-ray emission from supernova remnants following \cite{2010MNRAS.403L..16C}. 
To emphasize the effects of local sources, we chose a relatively high normalization for the local source spectrum corresponding to a distance of 10 proper kpc from a solar-metallicity galaxy with a star formation rate of 100 M$_{\odot}$ yr$^{-1}$. 
We assume that a fraction 5\% of the mechanical energy from the starburst is converted into ISM heating and X-ray emission \citep[][]{2008A&A...483...71M}. 
We assume that all the X-rays escape the galaxy but an escape fraction of 5\% for the softer ionizing photons between 1 and 4 Ry. 
Figure \ref{f:local_xrays} shows the results of this test and compares them with calculations including only the FG09 background (left) and only the FG09 background but with normalization multiplied by a factor of ten (right). 
For surface brightnesses $>10^{-20}$ erg s$^{-1}$ cm$^{-2}$ arcsec$^{-2}$ potentially observable in deep and stacked observations with MUSE and KCWI, there is no significant difference between the different ionization models. 
At lower surface brightnesses, some differences are noticeable with the inclusion of a local ionizing source. In particular, the C III line is suppressed at large radii while the C IV line is enhanced. 
However, at those very low surface brightnesses, our test overestimates the magnitude of the effect since we do not model the $\propto R^{-2}$ drop off of the local flux with radius. 
The effects are seen also in the calculation in which we simply boosted the cosmic UV background, indicating that it is primarily due to the enhancement of relatively soft ionizing photons rather than X-rays from local supernova remnants.

\begin{figure*}
\begin{center}
\includegraphics[width=\textwidth]{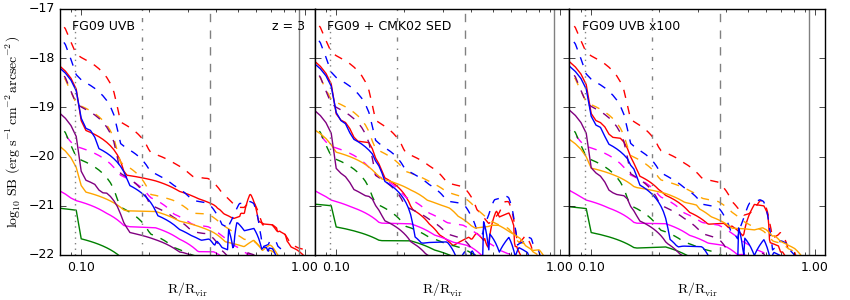}
\end{center}
\caption[]{
Test of the effects of a local source including X-rays on theoretical surface brightness profiles. We compare the median (solid) and mean (dashed) surface brightness profiles for five halos ({\bf z2h506}, {\bf z2h450}, {\bf z2h400}, {\bf z2h830}, and {\bf m12q}) at $z=3$. 
\emph{Left:} Profiles including only the FG09 cosmic ionizing background (default in main text). 
\emph{Middle:} Profiles including the sum of the FG09 background and a local source spectrum from \cite{2002A&A...392...19C}, which includes the contribution of supernova remnants to X-rays. The normalization of the X-ray contribution is set to the value expected at 10 proper kpc from a galaxy with a star formation rate of 100 M$_{\odot}$ yr$^{-1}$. \emph{Right:} Profiles including only the FG09 background but with normalization boosted by a factor of ten. 
The mean and median halo masses in $\log_{10}M_{\rm h}$ (M$_\odot$) are 11.46 and 11.44, respectively. 
The vertical dotted, dot-dashed, dashed and solid grey lines indicate radii of 5, 10, 20, and 50 proper kpc for the median mass. 
 }
\label{f:local_xrays} 
\end{figure*}
 
\end{document}